\documentclass[12pt]{article}
\pdfoutput=1
\usepackage[utf8]{inputenc}
\usepackage[normalem]{ulem}
\usepackage[english]{babel}%,frenchb
\usepackage{tabularx}
\usepackage{array}
\usepackage{graphics}
\usepackage[pdftex]{graphicx}
\usepackage{psfrag}
\usepackage{epsfig}
\usepackage{subfig}
\usepackage{mathtools}
\usepackage{amssymb}
\usepackage{bbold}
\usepackage{setspace}
\usepackage{rotating}
\usepackage{colortbl}
\usepackage{longtable}
\usepackage{slashed}
\usepackage{braket}
\usepackage{lineno}
\makeatletter
\usepackage{textcomp}
\usepackage[usenames,dvipsnames]{xcolor}
\usepackage{relsize}
\usepackage{cite}

\usepackage{float}

\usepackage{bbm}
\usepackage{bm}
\usepackage{enumerate}
\usepackage{amsmath}
\usepackage{verbatim}
\usepackage{hyperref}
\hypersetup{colorlinks,citecolor=nicegreen,linkcolor=niceblue}
\hypersetup{colorlinks=true}
\usepackage{cleveref}

\usepackage{pifont}% http://ctan.org/pkg/pifont

\allowdisplaybreaks

\setlongtables

\newcommand{\bea}{\begin{eqnarray}}
\newcommand{\eea}{\end{eqnarray}}
\newcommand{\beq}{\begin{equation}}
{
\newcommand{\eeq}{\end{equation}}
\newcommand{\ec}{\end{center}}
\newcommand{\bc}{\begin{center}}

\newcommand{\pdir}{p\kern -5.2pt\raise 0.2ex\hbox {/}}

\newcommand{\vdir}{v\kern -5.75pt\raise 0.15ex\hbox {/}}
\newcommand{\kdir}{k\kern -5.75pt\raise 0.15ex\hbox {/}}
\newcommand{\epsdir}{\epsilon\kern -5.0pt\raise 0.15ex\hbox {/}}
\newcommand{\bvdir}{\bar{v}\kern -5.75pt\raise 0.15ex\hbox {/}}
\newcommand{\Ddir}{D\kern -7.75pt\raise 0.20ex\hbox {/}}
\newcommand{\Adir}{A\kern -7.75pt\raise 0.20ex\hbox {/}}
\newcommand{\ldir}{l\kern -5.0pt\raise 0.2ex\hbox{/}}
\newcommand{\varepsdir}{\varepsilon\kern -5.5pt\raise 0.15ex\hbox{/}}

\newcommand{\mrm}[1]{\mathrm{#1}}

\newcommand{\re}[0]{\mrm{Re}}
\newcommand{\im}[0]{\mrm{Im}}
\makeatother

\definecolor{niceblue}{rgb}{0.15,0.15,0.6}
\definecolor{nicegreen}{rgb}{0.1,0.5,0.1}
\definecolor{Red}{rgb}{1.,0.,0.}

\definecolor{Green}{rgb}{0.2,.7,0.2}

\begin{document}
\unitlength = 1mm

\thispagestyle{empty} 
\begin{flushright}
\begin{tabular}{l}
%{\tt \footnotesize IJCLab-20-xx}\\
\end{tabular}
\end{flushright}
\begin{center}
\vskip 3.4cm\par
{\par\centering \textbf{\LARGE  
\Large \bf Looking for the effects of New Physics in the $\Lambda_b \longrightarrow \Lambda_c(\to \Lambda \pi) \ell \nu $  decay mode\\[0.3em] }}
\vskip 1.2cm\par
{\par\centering \large  
\sc Damir~Be\v{c}irevi\'c and Florentin Jaffredo}
{\par\centering \vskip 0.7 cm\par}
{\sl 
IJCLab, P\^ole Th\'eorie (B\^at. 210)\\
CNRS/IN2P3 et Universit\'e Paris-Saclay, 91405 Orsay, France. }
{\vskip 1.65cm\par}
\end{center}

\vskip 0.85cm
\begin{abstract}
One of the most pragmatic ways to look for the effects of New Physics at low energy scales is to study a detailed angular distribution of various decay modes, and in particular those based on $b\to c\ell\bar \nu_\ell$. In this paper we focus onto $\Lambda_b \longrightarrow \Lambda_c(\to \Lambda \pi) \ell \nu$ in a  generic effective theory setup in which, besides the Standard Model, we allow for all the possible covariant dimension-six effective 
operators capturing the contributions arising at high energy scales, beyond the Standard Model. We list a number of observables that could be used as a diagnostic tool to check for the presence of New Physics and to discriminate among its various scenarios. We also briefly comment on $\Lambda_c \to \Lambda l\nu$.
\end{abstract}
\newpage
\setcounter{page}{1}
\setcounter{footnote}{0}
\setcounter{equation}{0}
%%%%%%%%%%%%%%%%%%%%%%%%%%%%%%%%%%%%%%%%
\noindent

\renewcommand{\thefootnote}{\arabic{footnote}}
%\linenumbers

\setcounter{footnote}{0}

\tableofcontents

\newpage

%%%%%%%%%%%
%%%%%%%%%%%
%%%%%%%%%%%
%\section{Introduction}
%\label{sec:intro}

\section{Introduction}
Ever since the first indication of the lepton flavor universality violation (LFUV), reported by BaBar in Refs.~\cite{Lees:2012xj,Lees:2013uzd}, we witnessed a growing interest in the 
high energy physics community with a goal to clarify the situation and assess whether or not the LFUV is a real effect in the decay modes based on $b\to c\ell\bar \nu_\ell$ decays, with $\ell \in\{e,\mu, \tau\}$, mediated by the charged currents that occur at tree level 
in the Standard Model (SM). The BaBar collaboration was first to measure 
%%%%%%%%%%%%%%%%
\begin{equation}
R_{D^{(\ast)}} = \left. \dfrac{\mathcal{B}(B\to D^{(\ast)} \tau\bar{\nu})}{\mathcal{B}(B\to D^{(\ast)} l \bar{\nu})}\right|_{l\in \{e,\mu\}},
\label{eq:RD_definition}
\end{equation}
%%%%%%%%%%%%%%%%
and they found that both $R_{D}$ and $R_{D^{\ast}}$ are larger than predicted in the SM. Since $\mathcal{B}(B\to D^{(\ast)} l \bar{\nu})$ are known to be rather consistent with expectations, it has been inferred that $\mathcal{B}(B\to D^{(\ast)} \tau\bar{\nu})^\mathrm{exp}$ 
is larger than its SM prediction. To make that assessment clearer the lattice QCD community has been working to compute the relevant form factors so that the hadronic uncertainties 
could be minimized. While this has been achieved in the case of $B\to D \ell \bar{\nu}$ decays~\cite{Lattice:2015rga,Aoki:2019cca}, the first results regarding the $B\to D^\ast \ell \bar{\nu}_\ell$ decay need more clarification~\cite{FermilabLattice:2021cdg,Harrison:2021tol,Kaneko:2019vkx,DiCarlo:2021dzg}. On the experimental side, after combining significant contributions from various experiments~\cite{Huschle:2015rga, Aaij:2015yra,Hirose:2016wfn,Sato:2016svk,Belle:2019rba}, the HFLAV collaboration reported the following average values~\cite{Amhis:2022mac}:
\bea\label{eq:RDstar}
R_{D} =0.340(30) \,,\qquad 
R_{D^{\ast }} = 0.295(14)\,,
\eea 
which, together, are more than $3\sigma$ larger than evaluated in the SM. Another exclusive $b\to c\ell\bar \nu_\ell$ channel, in which the similar test of LFUV could be made, has been experimentally studied by the LHCb collaboration in Ref.~\cite{Aaij:2017tyk}, and the result  
%%%%%%%%%%%%%%%%
\begin{equation}
R_{J/\psi} =  \dfrac{\mathcal{B}(B_c\to J/\psi \tau\bar{\nu})}{\mathcal{B}(B\to J/\psi \mu \bar{\nu} )} = 0.71\pm 0.25\,,
\label{eq:Rpsi_definition}
\end{equation}
%%%%%%%%%%%%%%%%
again appears to be a little less than $2\sigma$ larger than its SM value~\cite{Harrison:2020nrv}.

The above observations have motivated many physicists to build scenarios that go beyond the SM (BSM) in order to accommodate the effects of LFUV while  keeping a large plethora of other processes compatible both with the SM and with experiment. Clearly, while the LFUV ratios $R_{D^{(\ast)}}$ and $R_{J/\psi}$ provide us with valuable information, they alone are insufficient to select among various possible BSM contributions to $b\to c\tau\bar \nu$. Indeed, much more information about the effects of physics BSM can be extracted from the angular distributions of the above-mentioned decay modes~\cite{BDetstar,Celis:2016azn}, some of which will be possible to study in the years to come,  at the Belle~II  and the LHC experiments.

Another exclusive channel, which is yet to be experimentally explored in the case of heavy lepton in the final state, is $\Lambda_b \to  \Lambda_c \ell \bar \nu $. So far the LHCb collaboration presented the results concerning the $q^2$-shape of the differential decay rate, $d\Gamma(\Lambda_b \to  \Lambda_c \mu \bar \nu )/dq^2$~\cite{Aaij:2017svr}, up to an unknown normalization factor. This year, LHCb reported the first measurement of $R_{\Lambda_c}$ and found~\cite{LHCb:2022piu}
\bea\label{eq:RLc}
R_{\Lambda_c} =  \dfrac{\mathcal{B}(\Lambda_b\to \Lambda_c \tau\bar{\nu})}{\mathcal{B}(\Lambda_b\to \Lambda_c \mu \bar{\nu} )} = 0.242\pm 0.076\,,
\eea
which is consistent with the SM prediction~\cite{Detmold:2015aaa}. That measurement can be improved in multiple ways. In this paper we will provide the expressions for the full angular distribution of this decay, including the subsequent decay $\Lambda_c\to \Lambda \pi$. We will then combine various coefficients to construct the observables which could provide us with valuable information concerning the BSM physics. While deriving the relevant expressions, we separately show the results for spin {\it up} and spin {\it down} of the outgoing lepton and/or baryon. In such a way we could propose new quantities, including the well known lepton polarization asymmetry. To illustrate the power of measuring angular observables relevant to $\Lambda_b \to  \Lambda_c \tau \bar \nu $, we monitor their integrated characteristics in several scenarios in which the BSM couplings are required to be consistent with $R_{D^{(\ast)}}^\mathrm{exp}$. Since the uncertainty of $R_{J/\psi}^\mathrm{exp}$ is large the BSM couplings selected from compatibility with $R_{D^{(\ast)}}^\mathrm{exp}$ are automatically consistent with $R_{J/\psi}^\mathrm{exp}$ as well.  We should also mention the constraints on the effective New Physics (NP) couplings arising from the LHC studies of the high-$p_T$ tails of the $pp \to \ell\nu$ differential cross section. Such constraints in this case are not yet competitive with those obtained from the low-energy observables, but in the future they might play ever more important role, cf. Refs.~\cite{Jaffredo:2021ymt}.

The reminder of this paper is organized as follows. In Sec.~\ref{sec:eft}, we remind the reader of the low-energy effective theory description of $b \to c \ell \bar\nu$ transitions, of the relevant hadronic matrix elements and write down the decay amplitude. In Sec.~\ref{sec:distr} we derive the detailed decay distribution both for  
 $\Lambda_b \to \Lambda_c \ell \nu $ and for $\Lambda_b \longrightarrow \Lambda_c (\to \Lambda \pi) \ell \nu $.  In Sec.~\ref{sec:obs} we discuss the observables that can be built from the full angular distribution, which we then integrate and evaluate in the SM in Sec.~\ref{sec:obs2}. In the same Section we give the ready-to-use formula  
for $R_{\Lambda_c}$, expressed in terms of the NP couplings. In Sec.~\ref{sec:pheno} we discuss the phenomenology and show how the angular observables can be used to validate or refute some of the BSM scenarios, commonly used to accommodate the deviations of $R_D^\mathrm{exp}$ and $R_{D^\ast}^\mathrm{exp}$ with respect to their SM values. We then summarize and conclude in Sec.~\ref{sec:concl}.

\section{Effective Theory, Matrix Elements, Decay Amplitude \label{sec:eft}}

To account for both the SM and the effects of physics BSM, we describe the $b\to c\ell \bar \nu$ process by the following low-energy effective field theory:
	\begin{align}
		\label{eq:lagrangian-lep-semilep}
		\mathcal{L}_{\mathrm{eff}} = & -2\sqrt{2}G_F V_{c b} \Big{[}(1+g_{V_L})\,(\overline{c}_{L}\gamma_\mu {b}_{L}) (\overline{\ell}_L\gamma^\mu\nu_L) + g_{V_R}\,(\overline{c}_{R}\gamma_\mu {b}_{R}) (\overline{\ell}_L\gamma^\mu\nu_L) \nonumber \\[0.3em]
		&+g_{S_L}(\mu)\,(\overline{c}_{R} b_{L})(\overline{\ell}_R \nu_L)+g_{S_R}(\mu)\,(\overline{c}_{L} b_{R})(\overline{\ell}_R \nu_L)+g_T(\mu)\,(\overline{c}_R \sigma_{\mu\nu}b_L)(\overline{\ell}_R \sigma^{\mu\nu} \nu_L)\Big{]}+\mathrm{h.c.}\,, \nonumber \\
		=& -\frac{G_F}{\sqrt{2}} V_{c b} \Big{[}(1+g_{V})\,(\overline{c}\gamma_\mu {b}) \bigl(\overline{\ell}\gamma^\mu (1-\gamma_5) \nu\bigr) - (1- g_A)\,(\overline{c}\gamma_\mu \gamma_5{b} ) \bigl(\overline{\ell}\gamma^\mu (1-\gamma_5)\nu\bigr)\nonumber \\[0.3em]
		&\qquad +g_S(\mu)\,(\overline{c} b)\bigl(\overline{\ell} (1-\gamma_5)\nu\bigr)+g_{P}(\mu)\,(\overline{c} \gamma_5 b)\bigl(\overline{\ell} (1-\gamma_5)\nu \bigr) \nonumber\\
		&\qquad + g_T(\mu)\,\bigl(\overline{c} \sigma_{\mu\nu}(1-\gamma_5)b\bigr)\bigl(\overline{\ell} \sigma^{\mu\nu} (1-\gamma_5)\nu \bigr)\Big{]}+\mathrm{h.c.}\,,
	\end{align} 
written in both commonly used bases of operators. The two sets of the NP couplings are related via, $g_{V,A}=g_{V_R}\pm g_{V_L}$, $g_{S,P}=g_{S_R}\pm g_{S_L}$. After setting all of the NP couplings to zero one obviously retrieves the SM Fermi theory. 

The main stumbling point in the discussion of the weak interaction processes of hadrons is the theoretical treatment of hadronic uncertainties. For the $\Lambda_b \to  \Lambda_c \ell \bar \nu $ decay, however, all of the relevant form factors have already been computed on the lattice~\cite{Detmold:2015aaa,Datta:2017aue}. In this paper we will use the same decomposition of the hadronic matrix elements as in the papers in which the form factors have been computed, namely:
\begin{align}
\label{eq:mels}
\langle  \Lambda_c \vert \overline{c}\gamma^\mu b \vert \Lambda_b \rangle &= \overline{u}_{\Lambda_c}\Bigg[ F_0(q^2)(M_{\Lambda_b}-M_{\Lambda_c})\frac{q^\mu}{q^2} + F_\perp(q^2)\left(\gamma^\mu-\frac{2M_{\Lambda_c}}{Q_+}p^\mu-\frac{2M_{\Lambda_b}}{Q_+}k^\mu\right) \nonumber \\ 
&\hspace{10mm}+ F_+(q^2)\frac{M_{\Lambda_b}+M_{\Lambda_c}}{Q_+}\left(p^\mu+k^\mu-(M_{\Lambda_b}^2-M_{\Lambda_c}^2)\frac{q^\mu}{q^2}\right) \Bigg] u_{\Lambda_b},\\
\label{eq:mels1}
\langle \Lambda_c \vert \overline{c}\gamma^\mu\gamma_5b \vert\Lambda_b \rangle &= -\overline{u}_{\Lambda_c}\Bigg[ G_0(q^2)(M_{\Lambda_b}+M_{\Lambda_c})\frac{q^\mu}{q^2} + G_\perp(q^2)\left(\gamma^\mu-\frac{2M_{\Lambda_c}}{Q_-}p^\mu-\frac{2M_{\Lambda_b}}{Q_-}k^\mu\right)\nonumber \\
&\hspace{10mm}+ G_+(q^2)\frac{M_{\Lambda_b}-M_{\Lambda_c}}{Q_-}\left(p^\mu+k^\mu-(M_{\Lambda_b}^2-M_{\Lambda_c}^2)\frac{q^\mu}{q^2}\right) \Bigg] u_{\Lambda_b},
\end{align}
which, by virtue of the vector and axial Ward identities, imply:
\begin{align}\label{eq:JS}
\langle  \Lambda_c \vert \overline{c}b  \vert  \Lambda_b \rangle &= F_0(q^2)\frac{M_{\Lambda_b}-M_{\Lambda_c}}{m_b-m_c}\overline{u}_{\Lambda_c}u_{\Lambda_b},\\
\label{eq:JP}
\langle  \Lambda_c \vert \overline{c}\gamma_5b \vert \Lambda_b \rangle &= G_0(q^2)\frac{M_{\Lambda_b}+M_{\Lambda_c}}{m_b+m_c}\overline{u}_{\Lambda_c}\gamma_5u_{\Lambda_b}.
\end{align}
Regarding the matrix element of the tensor density, 
\begin{align}
\label{eq:mels2}
\langle  \Lambda_c \vert \overline{c}i\sigma^{\mu\nu}b \vert \Lambda_b \rangle &= -\overline{u}_{\Lambda_c}\Bigg[
2 h_+(q^2)\frac{p^\mu k^\nu - p^\nu k^\mu}{Q_+}\nonumber\\
&\mkern-90mu+h_\perp(q^2)\left(\frac{M_{\Lambda_b}+M_{\Lambda_c}}{q^2}(q^\mu\gamma^\nu-q^\nu\gamma^\mu)-2\left(\frac{1}{q^2}+\frac{1}{Q_+}\right)(p^\mu k^\nu - p^\nu k^\mu)\right)\nonumber 
\\ 
&\mkern-90mu+\widetilde{h}_+(q^2)\left(i\sigma^{\mu\nu}-\frac{2}{Q_-}\left(M_{\Lambda_b}(k^\mu\gamma^\nu-k^\nu\gamma^\mu)-M_{\Lambda_c}(p^\mu\gamma^\nu-p^\nu\gamma^\mu)+p^\mu k^\nu-p^\nu k^\mu\right)\right)\\\nonumber
&\mkern-90mu+\widetilde{h}_\perp(q^2)\frac{M_{\Lambda_b}-M_{\Lambda_c}}{q^2Q_-}\bigg(\left(M_{\Lambda_b}^2-M_{\Lambda_c}^2-q^2\right)(\gamma^\mu p^\nu-\gamma^\nu p^\mu)\\\nonumber
&\mkern-90mu-\left(M_{\Lambda_b}^2-M_{\Lambda_c}^2+q^2\right)(\gamma^\mu k^\nu-\gamma^\nu k^\mu)+2(M_{\Lambda_b}-M_{\Lambda_c})(p^\mu k^\nu - p^\nu k^\mu)\bigg)\Bigg] u_{\Lambda_b},
\end{align}
from which one can also obtain $\langle  \Lambda_c \vert \overline{c}i\sigma^{\mu\nu}\gamma_5 b \vert \Lambda_b \rangle$ by simply using the relation
\begin{align}
\sigma^{\mu\nu}\gamma_5 = -\frac{i}{2}\epsilon^{\mu\nu\alpha\beta}\sigma_{\alpha\beta},
\end{align}
with the convention $\epsilon_{0123}=+1$. In the above decomposition of the matrix elements, $p$ and $k$ are the four-momenta of $\Lambda_b$ and $\Lambda_c$, respectively, while $q^2=(p-k)^2$, and $Q_\pm=(M_{\Lambda_b}\pm M_{\Lambda_c})^2 - q^2$. Kinematics and the explicit forms of spinors in the convenient reference frames are specified in Appendix. The polarization of the virtual vector boson, $\eta^\mu(\lambda)$ satisfies the completeness relation:
\begin{align}
\sum_{\lambda\in\{\pm,0,t\}} \eta^{\ast\mu}(\lambda)\eta^\nu(\lambda)\delta_\lambda = g^{\mu\nu}, \qquad \delta_{0}=- \delta_{\pm, t} = 1\,.
\end{align}

With all of the above ingredients in hands we can write the $\Lambda_b \to \Lambda_c  \ell \bar \nu $ amplitude as:
\begin{align}\label{eq:amplitude}
\frac{\mathcal{M}^{\lambda_b}_{\lambda_c \lambda_\ell }}{G_F V_{cb}/\sqrt{2}}=& H^{{\rm S-P},\lambda_b}_{\lambda_c}L^{\rm S-P}_{\lambda_\ell}+\sum_\lambda \delta_\lambda H^{{\rm V-A},\lambda_b}_{\lambda_c\lambda}L^{{\rm V-A}}_{\lambda_\ell\lambda}+\sum_{\lambda, \lambda'} \delta_\lambda\delta_{\lambda'}H^{{\rm T-T5},\lambda_b}_{\lambda_c\lambda\lambda'}L^{\rm T-T5}_{\lambda_\ell\lambda\lambda'},
\end{align}
where $\lambda_b$, $\lambda_c$, $\lambda_\ell$ and $\lambda^{(\prime )}$ are the polarization states of $\Lambda_b$, $\Lambda_c$, the outgoing lepton and the virtual vector boson, respectively.
The hadronic parts in the above decomposition are evaluated by using the explicit expressions for spinors, cf. Appendix. We get:
\begin{equation}
\setlength{\jot}{12pt}
\begin{aligned}
H^{{\rm S-P},\lambda_b}_{\lambda_c} &= g_S(\mu) \, \langle  \Lambda_c\vert\overline{c}b\vert \Lambda_b\rangle + g_P(\mu) \, \langle \Lambda_c\vert \overline{c}\gamma_5b\vert \Lambda_b\rangle, \\  
H^{{\rm V-A},\lambda_b}_{\lambda_c\lambda} &= (1+g_V)\,\eta^\ast_\mu(\lambda)\, \langle \Lambda_c\vert \overline{c}\gamma^\mu b\vert \Lambda_b \rangle - (1-g_A)\,\eta^\ast_\mu(\lambda)\, \langle \Lambda_c\vert \overline{c}\gamma^\mu\gamma_5b\vert \Lambda_b\rangle,\\
H^{{\rm T-T5},\lambda_b}_{\lambda_c\lambda\lambda'}&= g_T(\mu)\, \eta^\ast_\mu(\lambda)\eta^\ast_\mu(\lambda')\, \langle \Lambda_c\vert \overline{c}\sigma^{\mu\nu} b\vert \Lambda_b\rangle -g_T(\mu)\, \eta^\ast_\mu(\lambda)\eta^\ast_\mu(\lambda')\,\langle\Lambda_c\vert \overline{c}\sigma^{\mu\nu}\gamma_5 b\vert \Lambda_b\rangle .
\end{aligned}
\end{equation}
In a more explicit form, after inserting the hadronic matrix elements listed in Eq.~(\ref{eq:mels}--\ref{eq:mels2}) in the above expressions, the only non-zero components are the following ones:
\begin{equation}
\setlength{\jot}{12pt}
\begin{aligned}
H^{{\rm S-P},\pm}_{\pm} &= g_S \, \sqrt{Q_+}\frac{M_{\Lambda_b}-M_{\Lambda_c}}{m_b-m_c}\, F_0(q^2)  \mp  g_P\, \sqrt{Q_-}\frac{M_{\Lambda_b}+M_{\Lambda_c}}{m_b+m_c}\, G_0(q^2) ,\\
H^{{\rm V-A},\pm}_{\pm 0} &= (1+g_V)\, (M_{\Lambda_b}+M_{\Lambda_c})\frac{\sqrt{Q_-}}{\sqrt{q^2}}\, F_+(q^2) \mp (1-g_A)\, (M_{\Lambda_b}-M_{\Lambda_c})\frac{\sqrt{Q_+}}{\sqrt{q^2}}\, G_+(q^2),\\
H^{{\rm V-A},\pm }_{\pm t} &= (1+g_V)\, (M_{\Lambda_b}-M_{\Lambda_c})\frac{\sqrt{Q_+}}{\sqrt{q^2}}\, F_0(q^2) \mp (1-g_A)\, (M_{\Lambda_b}+M_{\Lambda_c})\frac{\sqrt{Q_-}}{\sqrt{q^2}}\, G_0(q^2),\\
H^{{\rm V-A},\mp}_{\pm \pm} &= -(1+g_V)\,\sqrt{2Q_-}\, F_\perp(q^2) \pm (1-g_A)\, \sqrt{2Q_+}\, G_\perp (q^2),\\ 
H^{{\rm T-T5},\pm}_{\pm +-} &=- g_T\, \left[ \sqrt{{Q_-}}\, {h_+}(q^2) \pm  \sqrt{{Q_+}}\, \widetilde{h}_+(q^2) \right] ,\\
H^{{\rm T-T5},\pm}_{\pm t0} &= g_T \, \left[  \sqrt{{Q_-}}\, {h_+}(q^2) \pm  \sqrt{{Q_+}}\, {\widetilde{h}_+}(q^2) \right] ,\\
H^{{\rm T-T5},\mp }_{\pm t\pm} &= -g_T \, \left[  (M_{\Lambda_b}+M_{\Lambda_c})\frac{\sqrt{2Q_-}}{\sqrt{q^2}} \, h_\perp(q^2) \pm (M_{\Lambda_b}-M_{\Lambda_c})\frac{\sqrt{2Q_+}}{\sqrt{q^2}}\, \widetilde{h}_\perp(q^2)\right],\\
H^{{\rm T-T5},\mp}_{\pm \pm 0} &= g_T\, \left[ \pm (M_{\Lambda_b}+M_{\Lambda_c})\frac{\sqrt{2Q_-}}{\sqrt{q^2}}\, h_\perp(q^2) + (M_{\Lambda_b}-M_{\Lambda_c})\frac{\sqrt{2Q_+}}{\sqrt{q^2}} \, \widetilde{h}_\perp(q^2)\right],
\end{aligned}
\end{equation}
where, for notational simplicity, we omit the renormalization scale dependence of the BSM couplings  $g_{S,P,T}$. In what follows we assume that scale to be $\mu = m_b$. 
Note also that $H^{{\rm T-T5},\lambda_b}_{\lambda_c \lambda\lambda'} = -H^{{\rm T-T5},\lambda_b}_{\lambda_c \lambda'\lambda}$.  
As for the leptonic parts,
\begin{equation}
\setlength{\jot}{10pt}
\begin{aligned}
L_{\lambda_\ell}^{\rm S-P} &= \langle\, \ell\overline{\nu}\, \vert \overline{u}(\lambda_\ell)v(\lambda_{\nu})\vert 0 \rangle , \\
L_{\lambda_\ell, \lambda}^{\rm V-A} &= \eta^\mu(\lambda)\, \langle\, \ell\overline{\nu}\, \vert \overline{u}(\lambda_\ell)\gamma_{\mu}v(\lambda_{\nu})\vert 0\rangle ,\\
L_{\lambda_\ell, \lambda\lambda'}^{\rm T-T5} &= i\eta^\mu(\lambda)\eta^\nu(\lambda') \langle\, \ell\overline{\nu}\,  \vert \overline{u}(\lambda_\ell)\sigma_{\mu\nu}v(\lambda_{\nu})\vert 0\rangle .
\end{aligned}
\end{equation}
The non-zero contributions read: 
\begin{equation}
\setlength{\jot}{10pt}
\begin{aligned}
L_{+}^{\rm S-P} &= 2\sqrt{q^2}\beta, &
L_{+, t}^{\rm V-A} &= 2\beta m_\ell, \\
L_{+, 0}^{\rm V-A} &= -2\beta m_\ell\cos\theta,&
L_{+, \pm}^{\rm V-A} &= \pm \sqrt{2}\beta m_\ell\sin\theta, \\
L_{-, 0}^{\rm V-A} &= 2\sqrt{q^2}\beta\sin\theta,&
L_{-, \pm}^{\rm V-A} &= \sqrt{2q^2}\beta(\pm \cos\theta+1), \\
L_{+,0\pm}^{\rm T-T5} &= -\sqrt{2q^2}\beta\sin\theta,&
L_{+, 0t}^{\rm T-T5} &=2\sqrt{q^2}\beta\cos\theta, \\
L_{+, -+}^{\rm T-T5} &=2\sqrt{q^2}\beta\cos\theta,&
L_{+, \pm t}^{\rm T-T5} &=\mp\sqrt{2q^2}\beta\sin\theta, \\
L_{-, 0\pm}^{\rm T-T5} &=-\sqrt{2}\beta m_\ell(\cos\theta\pm 1),&
L_{-, 0t}^{\rm T-T5} &=-2\beta m_\ell \sin\theta, \\
L_{-, -+}^{\rm T-T5} &=-2\beta m_\ell \sin\theta,&
L_{-, \pm t}^{\rm T-T5} &=\sqrt{2}\beta m_\ell(\mp \cos\theta-1),
\end{aligned}
\end{equation}
where $\beta = \sqrt{1- m_\ell^2/q^2}$. Similarly to the hadronic parts, also here $L_{\lambda_\ell, \lambda\lambda'}^{\rm T-T5} =-L_{\lambda_\ell, \lambda'\lambda}^{\rm T-T5}$. 
Moreover, we find that the leptonic amplitudes satisfy the following relations:
\begin{align}\label{eq:keyL}
L^{\lambda_\ell}_{\rm S-P} &= \frac{\sqrt{q^2}}{m_\ell}L^{\lambda_\ell}_{\rm V-A, t},\qquad L^{+1/2, 0t}_{\rm T-T5} = -\frac{\sqrt{q^2}}{m_\ell}L^{+1/2,0}_{\rm V-A},\qquad
L^{-1/2, 0t}_{\rm T-T5} = -\frac{m_\ell}{\sqrt{q^2}}L^{-1/2,0}_{\rm V-A}.
\end{align}
In other words, all of the leptonic amplitudes are proportional to $L_{\lambda_\ell, \lambda}^{\rm V-A}$, and we can therefore redefine the hadronic contributions as
\begin{equation}
\setlength{\jot}{10pt}
\begin{aligned}
\label{eq:Htilde}
\widetilde{H}^{\lambda_b+}_{\lambda_c\pm}&= -H^{{\rm V-A},\lambda_b}_{\lambda_c\pm}+\frac{2\sqrt{q^2}}{m_l}\left(\pm H^{{\rm T-T5},\lambda_b}_{\lambda_c\pm 0}+H^{{\rm T-T5},\lambda_b}_{\lambda_c\pm t}\right), \\
\widetilde{H}^{\lambda_b+}_{\lambda_c0}&= -H^{{\rm V-A},\lambda_b}_{\lambda_c0}+\frac{2\sqrt{q^2}}{m_l}\left( H^{{\rm T-T5},\lambda_b}_{\lambda_c+-}+H^{{\rm T-T5},\lambda_b}_{\lambda_c0t}\right), \\
\widetilde{H}^{\lambda_b+}_{\lambda_ct}&= H^{{\rm V-A},\lambda_b}_{\lambda_ct}+\frac{\sqrt{q^2}}{m_l}H^{{\rm S-P},\lambda_b}_{\lambda_c}, \\
\widetilde{H}^{\lambda_b-}_{\lambda_c\pm}&= -H^{{\rm V-A},\lambda_b}_{\lambda_c\pm}+\frac{2m_l}{\sqrt{q^2}}\left(\pm H^{{\rm T-T5},\lambda_b}_{\lambda_c\pm 0}+H^{{\rm T-T5},\lambda_b}_{\lambda_c \pm t}\right), \\
\widetilde{H}^{\lambda_b-}_{\lambda_c0}&= -H^{{\rm V-A},\lambda_b}_{\lambda_c0}+\frac{2m_l}{\sqrt{q^2}}\left( H^{{\rm T-T5},\lambda_b}_{\lambda_c+-}+H^{{\rm T-T5},\lambda_b}_{\lambda_c0t}\right), \\
\widetilde{H}^{\lambda_b-}_{\lambda_ct}&= H^{{\rm V-A},\lambda_b}_{\lambda_ct}+\frac{m_l}{\sqrt{q^2}}H^{{\rm S-P},\lambda_b}_{\lambda_c} .
\end{aligned}
\end{equation}
The relations~\eqref{eq:keyL}, therefore, help drastically simplifying the expression for the full decay amplitude~\eqref{eq:amplitude}, which now becomes
\begin{align}\label{eq:simple1}
\mathcal{M}^{\lambda_b\lambda_\ell}_{\lambda_c} &=  \frac{G_F V_{cb}}{\sqrt{2}}\, \sum_{\lambda\in\{\pm,0,t\}}\widetilde{H}^{\lambda_b\lambda_\ell}_{\lambda_c\lambda}L_{\lambda_\ell \lambda}^\mathrm{V-A},
\end{align}
just like in the SM, except that the whole set of NP contribution is now collected in $\widetilde{H}^{\lambda_b\lambda_\ell}_{\lambda_c\lambda}$.
Of all of the $32$ terms, only the following $12$ are nonzero: 
\begin{align}
\mathcal{M}^{++}_{+}&=\frac{G_F V_{cb}}{\sqrt{2}}\; 2\beta m_\ell\left(\widetilde{H}^{++}_{+t}-\cos\theta\ \widetilde{H}^{++}_{+0}\right), &\quad 
\mathcal{M}^{++}_{-}&=-\frac{G_F V_{cb}}{\sqrt{2}}\; \sqrt{2}\beta m_\ell\sin\theta\ \widetilde{H}^{++}_{--},\nonumber \\
\mathcal{M}^{-+}_{+}&=\frac{G_F V_{cb}}{\sqrt{2}}\; \sqrt{2}\beta m_\ell\sin\theta\ \widetilde{H}^{-+}_{++}, &
\mathcal{M}^{-+}_{-}&=\frac{G_F V_{cb}}{\sqrt{2}}\; 2\beta m_\ell\left(\widetilde{H}^{-+}_{-t}-\cos\theta\ \widetilde{H}^{-+}_{-0}\right),\nonumber \\
\mathcal{M}^{+-}_{+}&=\frac{G_F V_{cb}}{\sqrt{2}}\; 2\sqrt{q^2}\beta\sin\theta\ \widetilde{H}^{+-}_{+0}, &
\mathcal{M}^{+-}_{-}&=-\frac{G_F V_{cb}}{\sqrt{2}}\; \sqrt{2q^2}\beta(1-\cos\theta)\ \widetilde{H}^{+-}_{--},\nonumber \\
\mathcal{M}^{--}_{+}&=\frac{G_F V_{cb}}{\sqrt{2}}\; \sqrt{2q^2}\beta(1+\cos\theta )\ \widetilde{H}^{--}_{++}, &
\mathcal{M}^{--}_{-}&=-\frac{G_F V_{cb}}{\sqrt{2}}\; 2\sqrt{q^2}\beta\sin\theta\ \widetilde{H}^{--}_{-0}.
\end{align}
This, to our knowledge, is a new result and represents the most compact way to express the full $\Lambda_b\to \Lambda_c \ell\nu$ decay amplitude in a generic BSM scenario.

\section{$\Lambda_b \longrightarrow \Lambda_c (\to \Lambda \pi) \ell \nu $\label{sec:distr}}

Using the above expressions we can now write the angular distribution of the $\Lambda_b \longrightarrow \Lambda_c \ell \nu $ decay, $\ell \in\{e,\mu, \tau\}$. In this Section we discuss such a distribution for various polarization states of the outgoing $\Lambda_c$ and $\ell$. This will allow us to introduce polarization asymmetries. We will then consider the subsequent decay of $\Lambda_c\to \Lambda \pi$ and give the expression for the full angular distribution of $\Lambda_b \longrightarrow \Lambda_c (\to \Lambda \pi) \ell \nu $, again separating the rates according to the polarization states of the outgoing $\ell$ and $\Lambda$.

\subsection{Detailed $\Lambda_b \to \Lambda_c  \ell \nu $ Decay Rate}

We average over the polarizations of the initial state ($\Lambda_b$) and write the decay rate for each combination of $\lambda_c$ and $\lambda_\ell$. After inspection, we see that each such a differential decay rate can be written as
\begin{align}\label{eq:rate1}
\frac{d^2\Gamma^{\lambda_\ell}_{\lambda_c}}{dq^2d\cos{\theta}} &= a^{\lambda_\ell}_{\lambda_c}(q^2) + b^{\lambda_\ell}_{\lambda_c}(q^2)\cos\theta+c^{\lambda_\ell}_{\lambda_c}(q^2)\cos^2\theta.
\end{align}
The full decay rate is then obviously obtained by summing over $\lambda_c$ and $\lambda_\ell$. The explicit expressions for the coefficients $a^{\lambda_\ell}_{\lambda_c}$, $b^{\lambda_\ell}_{\lambda_c}$ and $c^{\lambda_\ell}_{\lambda_c}$ are:
\begin{align}
a_+^+(q^2)&=\mathcal{N}m_\ell^2\left(2\left|\widetilde{H}^{++}_{+t}\right|^2+\left|\widetilde{H}^{-+}_{++}\right|^2\right),&
a_-^+(q^2)&=\mathcal{N}m_\ell^2\left(2\left|\widetilde{H}^{-+}_{-t}\right|^2+\left|\widetilde{H}^{++}_{--}\right|^2\right),\nonumber \\
a_+^-(q^2)&=\mathcal{N}q^2\left(2\left|\widetilde{H}^{+-}_{+0}\right|^2+\left|\widetilde{H}^{--}_{++}\right|^2\right),&
a_-^-(q^2)&=\mathcal{N}q^2\left(2\left|\widetilde{H}^{--}_{-0}\right|^2+\left|\widetilde{H}^{+-}_{--}\right|^2\right),\nonumber \\
b^+_+(q^2)&=-4\mathcal{N}m_\ell^2\, \re\left(\overline{\widetilde{H}^{++}_{+0}}\widetilde{H}^{++}_{+t}\right),&
b^+_-(q^2)&=-4\mathcal{N}m_\ell^2\, \re\left(\overline{\widetilde{H}^{-+}_{-0}}\widetilde{H}^{-+}_{-t}\right),\nonumber \\
b^-_+(q^2)&=2\mathcal{N}q^2\left|\widetilde{H}^{--}_{++}\right|^2,&
b^-_-(q^2)&=-2\mathcal{N}q^2\left|\widetilde{H}^{+-}_{--}\right|^2, \nonumber  \\
c_+^+(q^2)&=\mathcal{N}m_\ell^2\left(2\left|\widetilde{H}^{++}_{+0}\right|^2-\left|\widetilde{H}^{-+}_{++}\right|^2\right), &
c_-^+(q^2)&=\mathcal{N}m_\ell^2\left(2\left|\widetilde{H}^{-+}_{-0}\right|^2-\left|\widetilde{H}^{++}_{--}\right|^2\right),\nonumber \\
c_+^-(q^2)&=\mathcal{N}q^2\left(-2\left|\widetilde{H}^{+-}_{+0}\right|^2+\left|\widetilde{H}^{--}_{++}\right|^2\right),&
c_-^-(q^2)&=\mathcal{N}q^2\left(-2\left|\widetilde{H}^{--}_{-0}\right|^2+\left|\widetilde{H}^{+-}_{--}\right|^2\right),
\end{align}
where
\begin{align}
\mathcal{N} \equiv \mathcal{N} (q^2)&= \frac{G_F^2|V_{cb}|^2\sqrt{\lambda_{{\Lambda_b\Lambda_c}}\left(q^2\right)}}{1024\pi^3M_{\Lambda_b}^3}\left(1-\frac{m_\ell^2}{q^2}\right)^2,
\end{align}
and $\lambda_{{\Lambda_b\Lambda_c}}(q^2) = Q_+ Q_-$. 
Using the above expressions we are now able to write the polarization asymmetry with respect to the outgoing lepton $\ell$ and with respect to $\Lambda_c$, the observables which we will come back to in the next Section.

Before continuing, it is interesting to note that, based on the above formulas, we have: 
\begin{align}\label{eq:ba}
b^-_+(q^2) &= a^-_+(q^2) +c^-_+(q^2), & b^-_-(q^2) &= -a^-_-(q^2) - c^-_-(q^2).
\end{align}
Since there are $4$ angular distributions of the differential rate~\eqref{eq:rate1}, each with $3$ coefficients, it means that one could construct at most $12$ linearly independent observables. 
That number reduces to $10$, thanks to the identities in Eq.~\eqref{eq:ba}.
Notice also that the form~\eqref{eq:rate1} is similar to what one gets for the semileptonic decays of the pseudoscalar mesons, such as $B\to D\ell \nu$. 
It can be shown that Eq.~\eqref{eq:ba} also holds true for $B\to D\ell \nu$, with an extra condition that $b^-(q^2)=0$, which comes from the fact that $L^{V-A}_{-t}=0$.~\footnote{Note that in the case of $B\to D\ell \nu$ there is only one index, referring to the polarization state of the outgoing lepton.}

\subsection{Inclusion of $\Lambda_c \to \Lambda \pi$}

In experiments one reconstructs $\Lambda_c$ from its decay products. This is also opportunity for defining more observables. 
Here we focus on $ \Lambda_c \to \Lambda \pi$ with a charged pion in the final state, and work in the narrow width approximation in which we can use the Breit-Wigner distribution:
\begin{align}
BW(k^2) &= \frac{1}{k^2-M_{\Lambda_c}^2+iM_{\Lambda_c}\Gamma_{\Lambda_c}} \quad \Longrightarrow \quad 
|BW(k^2)|^2 \simeq \frac{\pi}{M_{\Lambda_c}\Gamma_{\Lambda_c}}\delta\left(k^2-M_{\Lambda_c}^2\right).
\end{align}
In that way we can decompose the $4$-body amplitude in terms of the 3-body ones as
\begin{align}
\mathcal{M}^{(4)\lambda_b\lambda_\ell}_{\lambda_\Lambda} &= \sum_{\lambda_c=\pm}\langle\Lambda^{\lambda_\Lambda}\pi \vert \Lambda_c^{\lambda_c}\rangle \, \mathcal{M}^{\lambda_b\lambda_\ell}_{\lambda_c}\ BW(k^2).
\end{align}
A convenient parametrization of the matrix element $\langle\Lambda^{\lambda_\Lambda}\pi \vert \Lambda_c^{\lambda_c}\rangle$ is,
\begin{align}
\braket{\Lambda^+\pi \vert \Lambda_c^+} &= h_+\cos\left(\frac{\theta_\Lambda}{2}\right), & \braket{\Lambda^+\pi\vert \Lambda_c^-} &= h_-e^{i\phi}\sin\left(\frac{\theta_\Lambda}{2}\right),\nonumber \\
\braket{\Lambda^-\pi\vert \Lambda_c^+} &= -h_+e^{-i\phi}\sin\left(\frac{\theta_\Lambda}{2}\right), &
\braket{\Lambda^-\pi \vert\Lambda_c^-} &= h_-\cos\left(\frac{\theta_\Lambda}{2}\right) .
\end{align}
where $\theta_\Lambda$ is the angle between the $z$-axis and the direction of flight of $\Lambda$ in the $\Lambda_c$ rest frame, while the parameters $h_+$ and $h_-$ can be extracted from the total decay rate $\Gamma_{\Lambda_c\rightarrow\Lambda\pi}$ and the $\Lambda_c$-polarization 
asymmetry $\alpha$, viz.
\begin{align}
\Gamma_{\Lambda_c\rightarrow\Lambda\pi} &= \Gamma^+_{\Lambda_c\rightarrow\Lambda\pi}+\Gamma^-_{\Lambda_c\rightarrow\Lambda\pi} = 
  \frac{\sqrt{\lambda_{\Lambda_c\Lambda\pi}}}{32\pi M_{\Lambda_c}^3}\biggl( |h_+|^2+|h_-|^2\biggr) ,
\nonumber \\
\alpha &=\frac{ \Gamma^+_{\Lambda_c\rightarrow\Lambda\pi}-\Gamma^-_{\Lambda_c\rightarrow\Lambda\pi} }{ \Gamma^+_{\Lambda_c\rightarrow\Lambda\pi}+\Gamma^-_{\Lambda_c\rightarrow\Lambda\pi} }=  \frac{|h_+|^2-|h_-|^2}{|h_+|^2+|h_-|^2},
\end{align}
where $\lambda_{\Lambda_c\Lambda\pi} = (M_{\Lambda_c}^2 - (M_\Lambda + m_\pi)^2)(M_{\Lambda_c}^2 - (M_\Lambda - m_\pi)^2)$. The measurement of $\alpha$ has been recently improved at BES~III~\cite{Ablikim:2019zwe},  and now its world average is $\alpha = -0.84(9)$~\cite{ParticleDataGroup:2020ssz}. 
The sum $ |h_+|^2+|h_-|^2$, is traded for a parameter $\kappa$,
\begin{align}
\kappa &= |h_+|^2+|h_-|^2 =\frac{32\pi M_{\Lambda_c}^3}{\sqrt{\lambda_{\Lambda_c\Lambda\pi}}}\Gamma_{\Lambda_c\rightarrow\Lambda\pi} ,
\end{align}
so that
\begin{equation}
\setlength{\jot}{10pt}
\begin{aligned}
|h_+|^2\cos^2\left(\frac{\theta_\Lambda}{2}\right)+|h_-|^2\sin^2\left(\frac{\theta_\Lambda}{2}\right)&=(1+\alpha\cos\theta_\Lambda)\frac{\kappa}{2},  \\
\label{eq:alpha_identity}
|h_+|^2\cos^2\left(\frac{\theta_\Lambda}{2}\right)-|h_-|^2\sin^2\left(\frac{\theta_\Lambda}{2}\right)&=(\alpha+\cos\theta_\Lambda)\frac{\kappa}{2}.
\end{aligned}
\end{equation}
Using $\kappa$ and $\alpha$, we can describe $\Lambda_c\to \Lambda \pi$ and write the full angular distribution for the decay $\Lambda_b \to \Lambda_c (\to \Lambda \pi ) \ell\bar \nu_\ell$, which now involves $3$ angles: $\theta$, $\theta_\Lambda$ and $\phi$, cf. Appendix.
Similarly to what we did in Eq.~\eqref{eq:simple1}, in order to simplify the expression for the decay rate, we write 
\begin{align}
 \mathcal{M}^{(4)\lambda_b\lambda_\ell}_{\lambda_\Lambda}  &=\frac{G_F V_{cb}}{ \sqrt{2}}\, \sum_{\lambda\in\{\pm,0,t\}}\widehat{H}^{\lambda_b\lambda_\ell}_{\lambda_\Lambda\lambda}L_{\lambda_\ell\lambda}^{\rm V-A}\ BW(k^2),
\end{align}
where $\widehat{H}^{\lambda_b\lambda_\ell}_{\lambda_\Lambda\lambda}$ are obtained from $\widetilde{H}^{\lambda_b\lambda_\ell}_{\lambda_c\lambda}$, given in Eq.~\eqref{eq:Htilde}, by
\begin{align}
\begin{pmatrix}
\widehat{H}^{\lambda_b\lambda_\ell}_{+\lambda}\\
\widehat{H}^{\lambda_b\lambda_\ell}_{-\lambda}
\end{pmatrix}&=
\begin{pmatrix}
h_+\cos\left(\frac{\theta_\Lambda}{2}\right)&-h_+e^{-i\phi}\left(\frac{\theta_\Lambda}{2}\right)\\
h_-e^{i\phi}\sin\left(\frac{\theta_\Lambda}{2}\right)&h_-\cos\left(\frac{\theta_\Lambda}{2}\right)
\end{pmatrix}
\begin{pmatrix}
\widetilde{H}^{\lambda_b\lambda_\ell}_{+\lambda}\\
\widetilde{H}^{\lambda_b\lambda_\ell}_{-\lambda}
\end{pmatrix}.
\label{eq:def_Hhat}
\end{align}
The resulting $8$ amplitudes read:
\begin{equation}
\setlength{\jot}{10pt}
\begin{aligned}
\mathcal{M}^{(4)++}_{\lambda_\Lambda} &=  BW(k^2)\sqrt{2}\beta m_\ell\left(\sqrt{2}\widehat{H}^{++}_{\lambda_\Lambda t}-\sqrt{2}\cos\theta\widehat{H}^{++}_{\lambda_\Lambda 0}-\sin\theta\widehat{H}^{++}_{\lambda_\Lambda -}\right),\\
\mathcal{M}^{(4)-+}_{\lambda_\Lambda} &=  BW(k^2)\sqrt{2}\beta m_\ell\left(\sqrt{2}\widehat{H}^{-+}_{\lambda_\Lambda t}-\sqrt{2}\cos\theta\widehat{H}^{-+}_{\lambda_\Lambda 0}+\sin\theta\widehat{H}^{-+}_{\lambda_\Lambda -}\right),\\
\mathcal{M}^{(4)+-}_{\lambda_\Lambda} &= BW(k^2)\sqrt{2q^2}\beta\left((1-\cos\theta)\widehat{H}^{+-}_{\lambda_\Lambda +}+\sqrt{2}\sin\theta\widehat{H}^{+-}_{\lambda_\Lambda 0}\right),\\
\mathcal{M}^{(4)--}_{\lambda_\Lambda} &=  BW(k^2)\sqrt{2q^2}\beta\left((1+\cos\theta)\widehat{H}^{--}_{\lambda_\Lambda +}+\sqrt{2}\sin\theta\widehat{H}^{--}_{\lambda_\Lambda 0}\right),
\end{aligned}
\end{equation}
where the superscript ``$(4)$" indicates that we deal with the $4$-body decay. By combining the vector defined in Eq.~\eqref{eq:def_Hhat} with its conjugate and by using Eq.~(\ref{eq:alpha_identity}) we obtain two useful 
expressions, namely,
\begin{equation}
\setlength{\jot}{10pt}
\begin{aligned}
\label{eq:identity_width}
\overline{\widehat{H}^{\lambda_b\lambda_\ell}_{+\lambda}}\widehat{H}^{\lambda_b\lambda_\ell}_{+\lambda'}+\overline{\widehat{H}^{\lambda_b\lambda_\ell}_{-\lambda}}\widehat{H}^{\lambda_b\lambda_\ell}_{-\lambda'} &= \frac{\kappa}{2}\left(\overline{\widetilde{H}^{\lambda_b\lambda_\ell}_{+\lambda}}\widetilde{H}^{\lambda_b\lambda_\ell}_{+\lambda'}+\overline{\widetilde{H}^{\lambda_b\lambda_\ell}_{-\lambda}}\widetilde{H}^{\lambda_b\lambda_\ell}_{-\lambda'}\right) \\ 
&+\frac{\alpha\kappa\cos\theta_\Lambda}{2}\left(\overline{\widetilde{H}^{\lambda_b\lambda_\ell}_{+\lambda}}\widetilde{H}^{\lambda_b\lambda_\ell}_{+\lambda'}-\overline{\widetilde{H}^{\lambda_b\lambda_\ell}_{-\lambda}}\widetilde{H}^{\lambda_b\lambda_\ell}_{-\lambda'}\right) \\
&-\frac{\alpha\kappa\sin\theta_\Lambda}{2}\left(e^{-i\phi}\overline{\widetilde{H}^{\lambda_b\lambda_\ell}_{+\lambda}}\widetilde{H}^{\lambda_b\lambda_\ell}_{-\lambda'}+e^{i\phi}\overline{\widetilde{H}^{\lambda_b\lambda_\ell}_{-\lambda}}\widetilde{H}^{\lambda_b\lambda_\ell}_{+\lambda'}\right), \\
\overline{\widehat{H}^{\lambda_b\lambda_\ell}_{+\lambda}}\widehat{H}^{\lambda_b\lambda_\ell}_{+\lambda'}-\overline{\widehat{H}^{\lambda_b\lambda_\ell}_{-\lambda}}\widehat{H}^{\lambda_b\lambda_\ell}_{-\lambda'} &= \frac{\alpha\kappa}{2}\left(\overline{\widetilde{H}^{\lambda_b\lambda_\ell}_{+\lambda}}\widetilde{H}^{\lambda_b\lambda_\ell}_{+\lambda'}+\overline{\widetilde{H}^{\lambda_b\lambda_\ell}_{-\lambda}}\widetilde{H}^{\lambda_b\lambda_\ell}_{-\lambda'}\right)\\ 
&+\frac{\kappa\cos\theta_\Lambda}{2}\left(\overline{\widetilde{H}^{\lambda_b\lambda_\ell}_{+\lambda}}\widetilde{H}^{\lambda_b\lambda_\ell}_{+\lambda'}-\overline{\widetilde{H}^{\lambda_b\lambda_\ell}_{-\lambda}}\widetilde{H}^{\lambda_b\lambda_\ell}_{-\lambda'}\right)\\ 
&-\frac{\kappa\sin\theta_\Lambda}{2}\left(e^{-i\phi}\overline{\widetilde{H}^{\lambda_b\lambda_\ell}_{+\lambda}}\widetilde{H}^{\lambda_b\lambda_\ell}_{-\lambda'}+e^{i\phi}\overline{\widetilde{H}^{\lambda_b\lambda_\ell}_{-\lambda}}\widetilde{H}^{\lambda_b\lambda_\ell}_{+\lambda'}\right).
\end{aligned}
\end{equation}
The first of the above formulas shows that the summation over the spin projections of $\Lambda$ allows us to trade $h_+$ and $h_-$ for the overall factors $\kappa$ and $\alpha\kappa$. 
The same holds true in the second formula which is useful when considering the polarization asymmetry with respect to $\Lambda$. 
Furthermore, since $\widetilde{H}^{\lambda_b\lambda_\ell}_{+\lambda}$ and $\widetilde{H}^{\lambda_b\lambda_\ell}_{-\lambda}$ cannot be non-zero at the same time, the last line in both of the above formulas must be zero if $\lambda = \lambda'$.

\vskip 7mm

The final missing ingredient is the $4$-body phase space which we write as
\begin{align}
\mathrm{dLIPS} = \frac{1}{64(2\pi)^6}\frac{\sqrt{\lambda_{\Lambda_b\Lambda_c}(q^2)}}{2M_{\Lambda_b}^2}\frac{\sqrt{\lambda_{\Lambda_c\Lambda\pi}}}{2M_{\Lambda_c}^2}\left(1-\frac{m_\ell^2}{q^2}\right)dq^2dk^2d\cos\theta d\cos\theta_\Lambda d\phi,
\end{align}
and the full angular distribution reads:
\begin{align}\label{eq:full4}
\frac{d^4\Gamma}{dq^2d\cos\theta d \cos\theta_\Lambda d\phi} &=
\frac{1}{64(2\pi)^6}\frac{\sqrt{\lambda_{\Lambda_b\Lambda_c}(q^2)}}{2M_{\Lambda_b}^3}\frac{\sqrt{\lambda_{\Lambda_c\Lambda\pi}}}{2M_{\Lambda_c}^2}\left(1-\frac{m_\ell^2}{q^2}\right)\frac{1}{4}\sum_{\lambda_\ell\lambda_b\lambda_\Lambda}\int dk^2 \left|\mathcal{M}^{(4)\lambda_b\lambda_\ell}_{\lambda_\Lambda}\right|^2.
\end{align}

Before closing this Section we need to emphasize that $\Lambda_c\to \Lambda \pi$ as the secondary decay is our choice. One could equally choose $\Lambda_c\to p K_S$, since both of them have nearly equal branching fractions, $\mathcal{B}(\Lambda_c\to \Lambda \pi )=1.30(7)\%$ and $\mathcal{B}(\Lambda_c\to p K_S )=1.59(8)\%$~\cite{ParticleDataGroup:2020ssz}. Note also that these branching fractions are of the same order as $\mathcal{B}(\Lambda_c\to p K^-\pi^+)= 6.28(32)\%$~\cite{ParticleDataGroup:2020ssz}, which is currently used as the preferred reconstruction channel of $\Lambda_c$ in the LHCb analyses~
\cite{Aaij:2017svr,LHCb:2022piu}.
Currently, $\Lambda_c\to \Lambda \pi$ than the other channels is more advantageous because its asymmetry parameter $\alpha$ has been experimentally determined more accurately than in the case of $\Lambda_c\to p K_S$. More precisely,  $\alpha^{\Lambda\pi}= -0.84(9)$,  $\alpha^{p K_S} = 0.2(5)$~\cite{ParticleDataGroup:2020ssz}.

When possible we were able to compare our expressions with those that are available in the literature~\cite{Datta:2017aue,Gutsche:2015mxa,Mu:2019bin,Hu:2020axt,Shivashankara:2015cta,Penalva:2019rgt,Boer:2019zmp} and we find an overall agreement.

\section{Angular distribution and observables\label{sec:obs}}

If we do not sum over $\lambda_\ell$ in Eq.~\eqref{eq:full4} we can use the formulas given in Eq.~(\ref{eq:identity_width}), which appear to be particularly useful when combining the coefficients of the angular distribution in order to define various observables. Concerning the spin projections of $\Lambda$ we can either sum over them or take their difference. In this way we arrive to the angular distribution of the full decay,
\begin{equation}
\setlength{\jot}{10pt}
\begin{aligned}
\label{eq:0}
\frac{d^4\Gamma^{\lambda_\ell}}{dq^2d\cos\theta d \cos\theta_\Lambda d\phi} &= A_1^{\lambda_\ell}+A_2^{\lambda_\ell}\cos\theta_\Lambda\\ 
&+\left(B_1^{\lambda_\ell}+B_2^{\lambda_\ell}\cos\theta_\Lambda\right)\cos\theta\\
&+\left(C_1^{\lambda_\ell}+C_2^{\lambda_\ell}\cos\theta_\Lambda\right)\cos^2\theta\\
&+\left(D_3^{\lambda_\ell}\sin\theta_\Lambda\cos\phi+D_4^{\lambda_\ell}\sin\theta_\Lambda\sin\phi\right)\sin\theta\\
&+\left(E_3^{\lambda_\ell}\sin\theta_\Lambda\cos\phi+E_4^{\lambda_\ell}\sin\theta_\Lambda\sin\phi\right)\sin\theta\cos\theta,
\end{aligned}
\end{equation}
and the angular distribution of the $\Lambda$-polarization asymmetry: 
\begin{equation}
\setlength{\jot}{10pt}
\begin{aligned}
\label{eq:0bis}
\frac{d^4\mathcal{A}_{\Lambda}}{dq^2d\cos\theta d \cos\theta_\Lambda d\phi} &= \widetilde{A}_1^{\lambda_\ell}+\widetilde{A}_2^{\lambda_\ell}\cos\theta_\Lambda\\
&+\left(\widetilde{B}_1^{\lambda_\ell}+\widetilde{B}_2^{\lambda_\ell}\cos\theta_\Lambda\right)\cos\theta\\
&+\left(\widetilde{C}_1^{\lambda_\ell}+\widetilde{C}_2^{\lambda_\ell}\cos\theta_\Lambda\right)\cos^2\theta\\
&+\left(\widetilde{D}_3^{\lambda_\ell}\sin\theta_\Lambda\cos\phi+\widetilde{D}_4^{\lambda_\ell}\sin\theta_\Lambda\sin\phi\right)\sin\theta\\ 
&+\left(\widetilde{E}_3^{\lambda_\ell}\sin\theta_\Lambda\cos\phi+\widetilde{E}_4^{\lambda_\ell}\sin\theta_\Lambda\sin\phi\right)\sin\theta\cos\theta.
\end{aligned}
\end{equation}
The $q^2$-dependent coefficients entering Eq.~\eqref{eq:0} read:
\begin{align}\label{eq:A}
\nonumber
A_1^+ &=\kappa\beta^2\mathcal{N'}m_\ell^2\left(\left|\widetilde{H}^{++}_{--}\right|^2+2\left|\widetilde{H}^{++}_{+t}\right|^2+\left|\widetilde{H}^{-+}_{++}\right|^2+2\left|\widetilde{H}^{-+}_{-t}\right|^2\right),\\\nonumber
A_2^+ &=\alpha\kappa\beta^2\mathcal{N'}m_\ell^2\left(-\left|\widetilde{H}^{++}_{--}\right|^2+2\left|\widetilde{H}^{++}_{+t}\right|^2+\left|\widetilde{H}^{-+}_{++}\right|^2-2\left|\widetilde{H}^{-+}_{-t}\right|^2\right),\\\nonumber
B_1^+ &=-4\kappa\beta^2\mathcal{N'}m_\ell^2\, \re\left(\overline{\widetilde{H}^{-+}_{-t}}\widetilde{H}^{-+}_{-0}+\overline{\widetilde{H}^{++}_{+t}}\widetilde{H}^{++}_{+0}\right),\\\nonumber
B_2^+ &=4\alpha\kappa\beta^2\mathcal{N'}m_\ell^2\, \re\left(\overline{\widetilde{H}^{-+}_{-t}}\widetilde{H}^{-+}_{-0}-\overline{\widetilde{H}^{++}_{+t}}\widetilde{H}^{++}_{+0}\right),\\\nonumber
C_1^+ &=\kappa\beta^2\mathcal{N'}m_\ell^2\left(-\left|\widetilde{H}^{++}_{--}\right|^2+2\left|\widetilde{H}^{++}_{+0}\right|^2-\left|\widetilde{H}^{-+}_{++}\right|^2+2\left|\widetilde{H}^{-+}_{-0}\right|^2\right),\\\nonumber
C_2^+ &=\alpha\kappa\beta^2\mathcal{N'}m_\ell^2\left(\left|\widetilde{H}^{++}_{--}\right|^2+2\left|\widetilde{H}^{++}_{+0}\right|^2-\left|\widetilde{H}^{-+}_{++}\right|^2-2\left|\widetilde{H}^{-+}_{-0}\right|^2\right),\\\nonumber
D_3^+ &=2\sqrt{2}\alpha\kappa\beta^2\mathcal{N'}m_\ell^2\, \re\left(\overline{\widetilde{H}^{++}_{+t}}\widetilde{H}^{++}_{--}-\overline{\widetilde{H}^{-+}_{++}}\widetilde{H}^{-+}_{-t}\right),\\\nonumber
D_4^+ &=2\sqrt{2}\alpha\kappa\beta^2\mathcal{N'}m_\ell^2\, \im\left(\overline{\widetilde{H}^{++}_{+t}}\widetilde{H}^{++}_{--}-\overline{\widetilde{H}^{-+}_{++}}\widetilde{H}^{-+}_{-t}\right),\\\nonumber
E_3^+ &=2\sqrt{2}\alpha\kappa\beta^2\mathcal{N'}m_\ell^2\, \re\left(\overline{\widetilde{H}^{-+}_{++}}\widetilde{H}^{-+}_{-0}-\overline{\widetilde{H}^{++}_{+0}}\widetilde{H}^{++}_{--}\right),\\\nonumber
E_4^+ &=2\sqrt{2}\alpha\kappa\beta^2\mathcal{N'}m_\ell^2\, \im\left(\overline{\widetilde{H}^{-+}_{++}}\widetilde{H}^{-+}_{-0}-\overline{\widetilde{H}^{++}_{+0}}\widetilde{H}^{++}_{--}\right),\\\nonumber
%\end{align}
%\begin{align}
A_1^- &=\kappa\beta^2\mathcal{N'}q^2\left(\left|\widetilde{H}^{--}_{++}\right|^2+2\left|\widetilde{H}^{--}_{-0}\right|^2+\left|\widetilde{H}^{+-}_{--}\right|^2+2\left|\widetilde{H}^{+-}_{+0}\right|^2\right),\\\nonumber
A_2^- &=\alpha\kappa\beta^2\mathcal{N'}q^2\left(\left|\widetilde{H}^{--}_{++}\right|^2-2\left|\widetilde{H}^{--}_{-0}\right|^2-\left|\widetilde{H}^{+-}_{--}\right|^2+2\left|\widetilde{H}^{+-}_{+0}\right|^2\right),\\\nonumber
B_1^- &=2\kappa\beta^2\mathcal{N'}q^2\left(\left|\widetilde{H}^{--}_{++}\right|^2-\left|\widetilde{H}^{+-}_{--}\right|^2\right),\\\nonumber
B_2^- &=2\alpha\kappa\beta^2\mathcal{N'}q^2\left(\left|\widetilde{H}^{--}_{++}\right|^2+\left|\widetilde{H}^{+-}_{--}\right|^2\right),\\\nonumber
C_1^- &=\kappa\beta^2\mathcal{N'}q^2\left(\left|\widetilde{H}^{--}_{++}\right|^2-2\left|\widetilde{H}^{--}_{-0}\right|^2+\left|\widetilde{H}^{+-}_{--}\right|^2-2\left|\widetilde{H}^{+-}_{+0}\right|^2\right),\\\nonumber
C_2^- &=\alpha\kappa\beta^2\mathcal{N'}q^2\left(\left|\widetilde{H}^{--}_{++}\right|^2+2\left|\widetilde{H}^{--}_{-0}\right|^2-\left|\widetilde{H}^{+-}_{--}\right|^2-2\left|\widetilde{H}^{+-}_{+0}\right|^2\right),\\\nonumber
D_3^- &=-2\sqrt{2}\alpha\kappa\beta^2\mathcal{N'}q^2\, \re\left(\overline{\widetilde{H}^{--}_{-0}}\widetilde{H}^{--}_{++}+\overline{\widetilde{H}^{+-}_{--}}\widetilde{H}^{+-}_{+0}\right),\\\nonumber
D_4^- &=2\sqrt{2}\alpha\kappa\beta^2\mathcal{N'}q^2\, \im\left(\overline{\widetilde{H}^{--}_{-0}}\widetilde{H}^{--}_{++}+\overline{\widetilde{H}^{+-}_{--}}\widetilde{H}^{+-}_{+0}\right),\\\nonumber
E_3^- &=2\sqrt{2}\alpha\kappa\beta^2\mathcal{N'}q^2\, \re\left(-\overline{\widetilde{H}^{--}_{-0}}\widetilde{H}^{--}_{++}+\overline{\widetilde{H}^{+-}_{--}}\widetilde{H}^{+-}_{+0}\right),\\
E_4^- &=2\sqrt{2}\alpha\kappa\beta^2\mathcal{N'}q^2\, \im\left(\overline{\widetilde{H}^{--}_{-0}}\widetilde{H}^{--}_{++}-\overline{\widetilde{H}^{+-}_{--}}\widetilde{H}^{+-}_{+0}\right).
\end{align}
Once again $\beta = \sqrt{1-m_\ell^2/q^2}$, and 
\begin{align}
\kappa\beta^2\mathcal{N'} &= \frac{G_F^2|V_{cb}|^2}{4096\pi^4}\frac{\sqrt{\lambda_{\Lambda_b\Lambda_c}(q^2)}}{M_{\Lambda_b}^3}\left(1-\frac{m_\ell^2}{q^2}\right)^2\ \mathcal{B}\left(\Lambda_c\rightarrow\Lambda\pi\right).
\end{align}
The coefficients entering the angular distribution of the $\Lambda$-polarization asymmetry, cf. Eq.~\eqref{eq:0bis}, are not independent. We find:
\begin{equation}
\setlength{\jot}{10pt}
\begin{aligned}
\label{eq:B}
\widetilde{A}_1^\pm &=\alpha A_1^\pm, \qquad& \widetilde{A}_2^\pm &=A_2^\pm/\alpha,\\
\widetilde{B}_1^\pm &=\alpha B_1^\pm, & \widetilde{B}_2^\pm &= B_2^\pm/\alpha, \\
\widetilde{C}_1^\pm &=\alpha C_1^\pm, & \widetilde{C}_2^\pm &=C_2^\pm/\alpha, \\
\widetilde{D}_3^\pm &=D_3^\pm/\alpha, & \widetilde{D}_4^\pm &=D_4^\pm/\alpha, \\
\widetilde{E}_3^\pm &=E_3^\pm/\alpha, & \widetilde{E}_4^\pm &=E_4^\pm/\alpha. 
\end{aligned}
\end{equation}
This means that measuring the asymmetry with respect to the polarization of the outgoing $\Lambda$ does not bring us any new information about the BSM physics with respect to Eq.~\eqref{eq:0}. If one of these quantities becomes experimentally accessible, it can be used as another determination of the $\alpha$ parameter, and then both options for the secondary decay, $\Lambda_c\to \Lambda \pi$ and $\Lambda_c\to p K_S$ would become equally interesting to study.

As we can see from the expressions for the differential decay width, for each polarization of the outgoing lepton~\eqref{eq:0} there are $10$ $q^2$-dependent coefficients corresponding to $\lambda_\ell =+1/2$, and $10$ coefficients corresponding to 
$\lambda_\ell=-1/2$. In the latter case, however, it is easy to see that not all the coefficients are linearly independent. Instead, one can express $B_{1,2}^-$ in terms of $A_{1,2}^-$ and $C_{1,2}^-$. If we normalize each coefficient by the full differential decay rate, we get a total of $18$ observables which we discuss in the next Section.

We should again emphasize that the above considerations are made by averaging the polarization states of the initial $\Lambda_b$. If it is possible to produce polarized $\Lambda_b$'s. the number of observables would obviously be larger.

\section{Observables \label{sec:obs2}}

We already counted the number of independent coefficients in the angular distributions. We found that the angular analysis of the $3$-body decay, $\Lambda_b \to \Lambda_c \ell \nu$, can yield $10$ independent observables, while a detailed angular analysis of the $4$-body decay, $\Lambda_b \to \Lambda_c (\to \Lambda \pi ) \ell \nu$, allows us to define $18$ observables. 
From now on, by $O_i^\pm$ we will denote one of the angular coefficients, $O_i^\pm \in\{A_{1,2}^\pm,B_{1,2}^\pm,C_{1,2}^\pm,D_{3,4}^\pm,E_{3,4}^\pm\}$, and define $O_i = O_i^+ +O_i^-$ and $O_i^{\mathcal{A}} = O_i^+-O_i^-$, where the superscript $\mathcal{A}$ indicates that we deal with asymmetry with respect to the lepton polarization.~\footnote{We reiterate that the separation of the amplitudes with respect to the polarization of $\Lambda$ does not lead to any new interesting physics information. Instead, the separation of the amplitudes with respect to the lepton polarization does provide us with new information.} 

\subsection{Integrated observables, $R_{\Lambda_c}$ and more\label{RLc-pheno}}

Each of the above-mentioned coefficients (observables), entering the full angular distribution of $\Lambda_b \longrightarrow \Lambda_c (\to \Lambda \pi) \ell\nu $, is a $q^2$-dependent function. To get the integrated cha\-rac\-teristic of each one of them, we will integrate over the available phase space, $q^2\in [m_\ell^2, (M_{\Lambda_b}-M_{\Lambda_c})^2]$, and normalize by the full decay width which is given by:
\begin{align}
\frac{d\Gamma (\Lambda_b \to \Lambda_c \ell \nu)}{dq^2} &= 8\pi\left(A_1+\frac{C_1}{3}\right).
\end{align}
In other words we define,
\begin{align}
\label{eq:intchar}
\langle {O}_i \rangle &= \frac{8\pi}{\Gamma}\int\displaylimits_{m_\ell^2}^{(M_{\Lambda_b}-M_{\Lambda_c})^2}\!\! O_i \ dq^2,
\end{align}
where the factor $8\pi$ is chosen for convenience so that $\langle{A}_1\rangle+\langle{C}_1\rangle/3=1$.

Furthermore, for the quantities that are non-zero in the SM, we also introduce the ratios, 
\begin{align}\label{RRR}
\mathcal{R}(O_i^{(\mathcal{A})}) &= \frac{\langle{O}_i^{(\mathcal{A})}\rangle}{\langle{O}_{i}^{(\mathcal{A})}\rangle^\mrm{SM}},
\end{align}
which is a convenient way to measure the deviation of any given observable with respect to its SM value. \\

Using the definition similar to Eq.~\eqref{eq:RLc}, we write
\begin{align}
R_{\Lambda_c} =& \left. \frac{ \mathcal{B}(\Lambda_b \to \Lambda_c \tau \nu) }{ \mathcal{B}(\Lambda_b \to \Lambda_c l \nu) }\right|_{l\in \{e,\mu\}}. 
\end{align}
and we follow a common practice to assume that the New Physics affects only the decay with $\tau$ in the final state, i.e. $g_{V,A,S,P,T}^\tau \neq 0$, while $g_{V,A,S,P,T}^{\mu,e} = 0$. The generic expression for  $R_{\Lambda_c}$ can be conveniently written in terms of the NP couplings, $g_{V,A,S,P,T}\equiv g_{V,A,S,P,T}^\tau$, and the ``magic numbers" $a_i$ as:
\begin{equation}
\setlength{\jot}{12pt}
\begin{aligned}
\label{eq:Rlambda}
R_{\Lambda_c} =  & a_\mrm{ S}\, |g_S|^2 + a_\mrm{ VS}\, {\re\left[(1+g_{\mrm V}) {g_{\mrm S}}^\ast\right]} + a_\mrm{ P}\, |g_P|^2  + a_\mrm{ AP} \, \re\left[(g_A-1) {g_P}^\ast\right] + a_\mrm{ V}  \,  |1+g_V|^2
 \nonumber \\
& + a_\mrm{ TV}\, \re\left[g_T (1+{g_V}^*)\right] + a_\mrm{ A}\, |1-g_A|^2  + a_\mrm{ TA}\, \re\left[g_T ({g_A}^\ast -1)\right]+a_\mrm{ T}\, |g_T|^2\, .
\end{aligned}
\end{equation}
Since the parameter $\alpha$ does not enter the expression for the decay rate, the errors on the values of $a_i$ are practically entirely due to form factors. We computed all of the magic numbers $a_i$ and their values are given in Tab.~\ref{tab:1}. The correlation matrix has the following form   
%%%%%%%%%%%%%%%%%%%%%
\begin{table}[t]
\renewcommand{\arraystretch}{1.5}
\centering
\scalebox{.9}{
\begin{tabular}{|c||c|c|c|c|c|c|c|c|c|}
\hline 
$a_i$ &$a_\mrm{S}$ & $a_\mrm{VS}$  & $a_\mrm{P}$ & $a_\mrm{AP}$ &  $a_\mrm{V}$ & $a_\mrm{TV}$& $a_\mrm{A}$& $a_\mrm{TA}$& $a_\mrm{T}$ \\\hline\hline
Central Value & $0.1011$&$0.1414$&$0.0105$&$-0.0272$&$0.1061$ &$0.2947$&$0.2270$&$1.3289$&$3.4734$\\  \hline
Error & $0.0048$&$0.0071$&$0.0005$&$0.0013$&$0.0055$ &$0.0156$&$0.0093$&$0.0528$&$0.1448$  \\ \hline
\end{tabular}
}
\caption{ \sl \small Central values and uncertainty on each of the nine magic numbers $a_i$ entering the expression for $R_{\Lambda_c}$ given in Eq.~\eqref{eq:Rlambda}. Correlation matrix is given in the text.}
\label{tab:1} 
\end{table}
%%%%%%%%%%%%%%%%%%%%
\[ 
\mrm{Corr}_a =
 \left(
\begin{array}{ccccccccc}
$1$&$0.993$&$0.6985$&$-0.7$&$0.9492$&$0.764$&$0.8137$&$0.816$&$0.7969$\\
 \cdot   &$1$&$0.6695$&$-0.6741$&$0.9666$&$0.7611$&$0.777$&$0.7784$&$0.7669$\\
 \cdot & \cdot &$1$&$-0.9987$&$0.7091$&$0.7203$&$0.8545$&$0.7893$&$0.7408$\\
 \cdot & \cdot & \cdot &$1$&$-0.7159$&$-0.7237$&$-0.8554$&$-0.7892$&$-0.7402$\\
 \cdot & \cdot & \cdot & \cdot &$1$&$0.8389$&$0.7555$&$0.7464$&$0.754$\\
 \cdot & \cdot & \cdot & \cdot & \cdot &$1$&$0.7229$&$0.7295$&$0.8344$\\
 \cdot & \cdot & \cdot & \cdot & \cdot & \cdot &$1$&$0.9703$&$0.877$\\
 \cdot & \cdot & \cdot & \cdot & \cdot & \cdot & \cdot &$1$&$0.9483$\\
 \cdot & \cdot & \cdot & \cdot & \cdot & \cdot & \cdot & \cdot &$1$\\
\end{array}
\right)  \;,
\]
where the order of rows and columns corresponds to the order of the magic numbers in Tab.~\ref{tab:1}. Obviously, if we set all of the NP couplings to zero we obtain:
\begin{align}
R_{\Lambda_c}^\mrm{SM} =  0.333\pm 0.013\, .
\end{align}
Before closing this Section we also give the Standard Model values for all $\langle{O}_{i}^{(\mathcal{A})}\rangle$, in the case of $\tau$ in the final state. We find:
\begin{equation}
\setlength{\jot}{10pt}
\begin{aligned}
&\langle A_1\rangle_\mrm{SM} =  1.035(1)\,, &&\langle A_2\rangle_\mrm{SM} =  0.658(6)\,,  & & \langle B_1 \rangle_\mrm{SM} =  0.049(8)\,, & &\langle B_2\rangle_\mrm{SM} =  -0.093(9)\,, \\  
&\langle C_1\rangle_\mrm{SM} =  -0.106(3)\,,   &&\langle C_2\rangle_\mrm{SM} =  -0.095(2)\,,  & &\langle D_3 \rangle_\mrm{SM} =  0.189(8) \,, & & \langle D_4 \rangle_\mrm{SM} =  0\,,  \\
&\langle E_3\rangle_\mrm{SM} =   0.069(2)\,,   &&\langle E_4\rangle_\mrm{SM} =  0\,,  & &  \, & & \,  
\end{aligned}
\end{equation}
and 
\begin{equation}
\setlength{\jot}{10pt}
\begin{aligned}
&\langle A_1^\mathcal{A}\rangle_\mrm{SM} =  -0.405(6)\,, & &\langle A_2^\mathcal{A}\rangle_\mrm{SM} =  -0.261(4)\,, & &\langle B_1^\mathcal{A}\rangle_\mrm{SM} =  0.667(6)\,, & &\langle B_2^\mathcal{A}\rangle_\mrm{SM} =  0.761(2)\,, \\
&\langle C_1^\mathcal{A}\rangle_\mrm{SM} =  0.293(7)\,, & &\langle C_2^\mathcal{A}\rangle_\mrm{SM} =  0.300(7)\,, & &\langle D_3^\mathcal{A}\rangle_\mrm{SM} =  -0.492(8)\,, & &\langle D_4^\mathcal{A}\rangle_\mrm{SM} =  0\,, \\
&\langle E_3^\mathcal{A}\rangle_\mrm{SM} =  -0.172(6)\,, & &\langle E_4^\mathcal{A} \rangle_\mrm{SM} =  0 \,. & &  
\end{aligned}
\end{equation}
We should add, once again, that due to the fact that $\alpha (A_1^- + C_1^-)=B_2^-$ and $(A_2^- + C_2^-)=\alpha B_1^-$, two of the above observables are not independent, so that the final number is indeed $18$, as discussed before. 

\section{Illustration and Phenomenology \label{sec:pheno}}

We made an extensive analysis of all of the observables mentioned so far, and found that the following $6$ exhibit more pronounced sensitivity to the presence of physics BSM:
\begin{itemize}
\item Ratio $R_{\Lambda_c}$, which in the SM is predicted to be $R_{\Lambda_c}^\mathrm{SM} = 0.333(13)$.
\item Forward-backward asymmetry, $\langle \mathcal{A}_{\mathrm{fb}}\rangle = \langle B_1\rangle/2$, which in the SM is expected to be $\langle \mathcal{A}_{\mathrm{fb}}\rangle^\mathrm{SM} = 0.049(8)$.
\item Lepton polarization asymmetry, $\langle\mathcal{A}_{\tau}\rangle= \langle A_1^\mathcal{A}\rangle + \langle C_1^\mathcal{A}\rangle/3$. In the SM, $\langle\mathcal{A}_{\tau}\rangle^\mathrm{SM} = -0.307(7)$.
\item Asymmetry ``$\pi/3$", $\langle\mathcal{A}_{\pi/3}\rangle= \langle C_1\rangle/4$, the SM value of which is $\langle\mathcal{A}_{\pi/3}\rangle^\mathrm{SM}=-0.027(1)$.~\footnote{From the full angular distribution~\eqref{eq:0}, after integrating over $q^2$, $\theta_\Lambda$ and $\phi$, $\mathcal{A}_{\pi/3}$ is obtained by selecting the events as follows: 
\begin{align}
\mathcal{A}_{\pi/3} & = \frac{1}{\Gamma}\left[ \int_0^{\pi/3} + \int_{2 \pi/3}^{\pi} - \int_{ \pi/3}^{2 \pi/3}\right] \frac{d\Gamma}{d \cos\theta} \sin\theta\, d\theta .
\end{align}
}
\item $\langle D_4\rangle$, which is strictly zero in the SM, $\langle D_4\rangle^\mathrm{SM} =0$.
\item $\langle E_4^\mathcal{A}\rangle$, also zero in the SM, $\langle E_4^\mathcal{A}\rangle^\mathrm{SM} =0$.
\end{itemize}
The last two quantities become non-zero only if the imaginary part of one of the NP couplings is different from zero.

We reiterate that we assume that the LFUV in the $b\to c\ell \bar \nu$ decays originates from the pronounced NP coupling to $\tau$, $g_{V,A,S,P,T}^{(\tau )}\neq 0$, while we keep $g_{V,A,S,P,T}^{(e,\mu)}=0$. 
In order to select the plausible values of the couplings $g_{V,A,S,P,T}\equiv g_{V,A,S,P,T}^{(\tau )}$ we use the current experimental values of $R_D$ and $R_{D^\ast}$, and extract the allowed ranges of each of the couplings by using the expressions presented in Ref.~\cite{Angelescu:2021lln}. 
Obviously, since we have only two experimental input values we cannot simultaneously vary all of the BSM couplings. Instead, we restrain our attention to four scenarios of physics BSM that have been actively investigated in recent years. More specifically, we either allow only $g_{V_L}$ or only $g_{S_L}$ to be non-zero, or 
we consider a peculiar combination of two BSM couplings which satisfy $g_{S_L}=\pm 4 g_T$ at the high energy scale, $\mu = \mathcal{O}(1\;\mathrm{TeV})$. Due to the renormalization group running,  at the $\mu = m_b$ this last relation becomes $ g_{S_L}\simeq  8.1  g_T$, and $ g_{S_L}\simeq  -8.5  g_T$~\cite{Gonzalez-Alonso:2017iyc}. 
In Tab.~\ref{tab:2} we present the best fit values for each of the scenarios. We first allow the couplings to be complex, and then impose to be either fully real or fully imaginary. Notice that in the scenario with 
$g_{S_L}=+ 4 g_T$ there is no real solution that would accommodate both $R_D^\mathrm{exp}$ and $R_{D^\ast}^\mathrm{exp}$~\cite{Angelescu:2021lln}.
%%%%%%%%%%%%%%%%%%%%%
\begin{table}[t]
\def\arraystretch{1.8}
\centering
\begin{tabular}{c|c|c|c|}
\cline{2-4}
                                      & $g_i\in \mathbb{C}$         & $\qquad g_i\in \mathbb{R}\qquad $ & $g_i\in \mathbb{C}$ and $\re[g_i] = 0$ \\ \hline
\multicolumn{1}{|c|}{$g_{V_L}$}       & $0.074$           & $0.074$       & $\pm 0.39\, i$       \\ \hline
\multicolumn{1}{|c|}{$g_{S_L}$}       & $-0.76\pm 0.80\, i$  & $0.12$        & $\pm 0.48\, i $       \\ \hline
\multicolumn{1}{|c|}{$g_T$}           & $0.10 \pm 0.17\, i$  & $-0.032$      & $\pm 0.10\, i$        \\ \hline
\multicolumn{1}{|c|}{$g_{S_L}=+8.1\, g_T$}  & $-0.094\pm 0.51\, i$ &      N.A.        & $\pm 0.48\, i$        \\ \hline
\multicolumn{1}{|c|}{$g_{S_L}=-8.5\, g_T$} & $0.16$            & $0.16$        & $\pm 0.48\, i$        \\ \hline
\end{tabular}
\caption{ \sl \small Best fit values for the couplings obtained by requiring $R_D$ and $R_{D^*}$ to be consistent with the experimental values. Apart from $g_{V_L}$, all the couplings are scale dependent. The above results refer to the scale $\mu=m_b$. Notice that the last two scenarios verify $g_{S_L}=\pm 4 g_T$ at $\mu \simeq 1\, \mathrm{TeV}$. }
\label{tab:2}
\end{table}
%%%%%%%%%%%%%%%%%%%%

For each of the values of the couplings given in Tab.~\ref{tab:2} we compute the observables mentioned above and compare them with their SM values, following Eq.~\eqref{RRR}. For the quantities which are zero in the SM, such as $\langle D_4\rangle$ and $\langle E_4^\mathcal{A}\rangle$, we just give the values which are non-zero in the presence of NP. The response of the observables mentioned above to the non-zero BSM 
couplings is given in Tab.~\ref{tab:obs_fit_points}.
%%%%%%%%%%%%%%%%%%%%
\begin{table}[H]
\def\arraystretch{2.1}
\centering
\begin{tabular}{ccllllll}
\hline
\multicolumn{2}{|c|}{Observable}                                                  & $\mathcal{R}( R_{\Lambda_c} )$ & $\mathcal{R}( \mathcal{A}_{\rm fb})$ & $\mathcal{R}( \mathcal{A}_{\tau})$ & $\mathcal{R}( \mathcal{A}_{\pi/3})$ & $\langle D_4\rangle$   & \multicolumn{1}{l|}{$\langle E_4^\mathcal{A}\rangle$ } \\ \hline 
\multicolumn{1}{|c|}{ $g_{V_L}$ }        & \multicolumn{1}{c|}{cplx} & 1.153(0)                                       & 1                                                       & 1                                                         & 1                                                          & 0       & \multicolumn{1}{l|}{0}               \\ \hline
\multicolumn{1}{|c|}{  $g_{S_L}$}          & \multicolumn{1}{c|}{cplx} & 1.147(3)                                       & -0.77(25)                                               & 0.45(1)                                                   & 0.87(0)                                                    & 0.12(0) & \multicolumn{1}{l|}{0}               \\
\multicolumn{1}{|c|}{}                                   & \multicolumn{1}{c|}{real} & 1.046(1)                                       & 1.24(4)                                                 & 0.81(1)                                                   & 0.96(0)                                                    & 0       & \multicolumn{1}{l|}{0}               \\
\multicolumn{1}{|c|}{}                                   & \multicolumn{1}{c|}{im} & 1.077(1)                                       & 0.93(0)                                                 & 0.70(1)                                                   & 0.93(0)                                                    & 0.08(0) & \multicolumn{1}{l|}{0}               \\ \hline
\multicolumn{1}{|c|}{ $g_{T}$}           & \multicolumn{1}{c|}{cplx} & 1.095(8)                                       & 2.92(39)                                                & 0.50(1)                                                   & 1.38(2)                                                    & 0.10(0) & \multicolumn{1}{l|}{-0.25(1)}        \\
\multicolumn{1}{|c|}{}                                   & \multicolumn{1}{c|}{real} & 1.110(2)                                       & 1.51(15)                                                & 0.99(0)                                                   & 0.91(0)                                                    & 0       & \multicolumn{1}{l|}{0}               \\
\multicolumn{1}{|c|}{}                                   & \multicolumn{1}{c|}{im} & 1.104(2)                                       & 1.83(17)                                                & 0.88(0)                                                   & 1.02(1)                                                    & 0.06(0) & \multicolumn{1}{l|}{-0.15(0)}        \\ \hline
\multicolumn{1}{|c|}{ $g_{S_L}=4g_{T}$}  & \multicolumn{1}{c|}{cplx} & 1.137(1)                                       & 2.02(22)                                                & 0.74(1)                                                   & 0.93(1)                                                    & 0.11(0) & \multicolumn{1}{l|}{-0.09(0)}        \\
\multicolumn{1}{|c|}{}                                   & \multicolumn{1}{c|}{im} & 1.114(1)                                       & 1.91(19)                                                & 0.66(1)                                                   & 0.94(0)                                                    & 0.11(0) & \multicolumn{1}{l|}{-0.08(0)}        \\ \hline
\multicolumn{1}{|c|}{ $g_{S_L}=-4g_{T}$}  & \multicolumn{1}{c|}{cplx} & 1.006(3)                                       & 1.17(32)                                                & 0.72(1)                                                   & 1.00(0)                                                    & 0       & \multicolumn{1}{l|}{0}               \\
\multicolumn{1}{|c|}{}                                   & \multicolumn{1}{c|}{im} & 1.114(1)                                       & 0.53(7)                                                 & 0.66(1)                                                   & 0.94(0)                                                    & 0.04(0) & \multicolumn{1}{l|}{0.09(0)}         \\ \hline
\end{tabular}
\caption{ \sl \small Illustration of the change of the observable with respect to its SM value for different choices of the New Physics couplings chosen as discussed in the text and explicitly given in Tab.~\ref{tab:2}. Notice that we separate the cases in which $g_i\in \mathbb{C}$, $g_i\in \mathbb{R}$ or purely imaginary. The error on $D_4$ and $E_4^\mathcal{A}$ only takes into account the error on the form factors, not on $\alpha$, which is taken to be $\alpha =0.82$~\cite{Boer:2019zmp}. 
\label{tab:obs_fit_points} }
\end{table}
%%%%%%%%%%%%%%%%%%%%
%%%%%%%%%%%%%%%%%%%%
Therefore if we measure the angular observables of $\Lambda_b \to \Lambda_c \tau \nu$ (without the secondary decay) we can already have access to four of the above-mentioned quantities, which in turn can help us in identifying the Lorentz structure of the BSM contribution.
To figure out whether or not there is a BSM phase, one has to include the secondary decay $\Lambda_b \to \Lambda_c(\to \Lambda \pi ) \tau \nu$  which then open a possibility to test if there is a non-zero contribution to  $\langle D_4\rangle$ and  $\langle E_4^\mathcal{A}\rangle$. 

We now go through various scenarios to further illustrate the usefulness of the quantities discussed in this Section. 

\subsection{Only $g_{V_L}\neq 0$}

\begin{figure}[t]
\centering    \includegraphics[scale=.5]{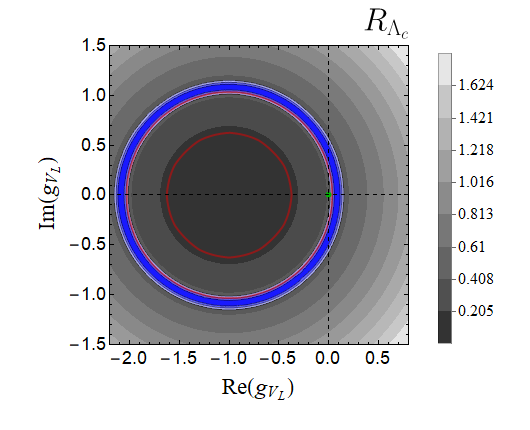}
        \caption{\sl  \small $R_{\Lambda_c}$ is plotted in the plane of $\re ( g_{V_L}) $ -  $\im ( g_{V_L}) $. Blue circle corresponds to the values of $ g_{V_L}$ allowed by simultaneously requiring $R_D$ and $R_{D^*}$ to be compatible with $R_D^\mathrm{exp}$ and $R_{D^*}^\mathrm{exp}$. We also show the $2\sigma$ region defined by $R_{\Lambda_c}^\mathrm{exp}$ (between the two red circles). The green dot corresponds to the SM value. Various values of $R_{\Lambda_c}$ are shown by the graded gray regions.}
    \label{fig:gVL}
\end{figure}

Considering the scenarios in which the only BSM coupling different from zero is $g_{V_L}$, practically all of the observables are SM-like, because 
$1+g_{V_L}$ enters as an overall factor with respect to the SM Lagrangian, cf. Eq.~\eqref{eq:lagrangian-lep-semilep}. Since the angular observables are normalized to the decay rate, the effect of this coupling is not changing the SM predictions. The only exception is precisely $R_{\Lambda_c}$, the value of which does change, as we show in Fig.~\ref{fig:gVL}. 
An example of the explicit BSM model that falls into this category is the one with a light [$\mathcal{O}(1\,\mathrm{TeV})$] vector leptoquark~\cite{vlq}, . 

\subsection{$g_{S_L, T}\neq 0$}

In the case in which only $g_{S_L}\neq 0$, we find that a particularly sensitive quantity is $\langle \mathcal{A}_{\pi/3}\rangle$ and the $\tau$-polarization asymmetry, $\langle \mathcal{A}_{\tau}\rangle$. However, since the 
preferred values of $g_{S_L}$ by $R_D^\mathrm{exp}$ and $R_{D^*}^\mathrm{exp}$ have a large imaginary part, also $\langle D_4\rangle$ may be significantly different from zero, cf. Tab.~\ref{tab:obs_fit_points}. A representative example of such a scenario would be the extension of the SM with two Higgs doublets (2HDM), which provides a new tree-level mediator for the $B\to D^{(\ast )} \tau \bar \nu_\tau$, namely the charged Higgs boson. The complex coupling, selected by $R_D^\mathrm{exp}$ and $R_{D^*}^\mathrm{exp}$, is however unusual and inconsistent with a Type~II 2HDM~\cite{Celis:2016azn,2hdm}. Note also that a purely real $g_{S_L}$ is inconsistent with $R_D^\mathrm{exp}$ and $R_{D^*}^\mathrm{exp}$ to almost $3\sigma$.  
\begin{figure}[t]
\hspace*{-11mm} \includegraphics[scale=.5]{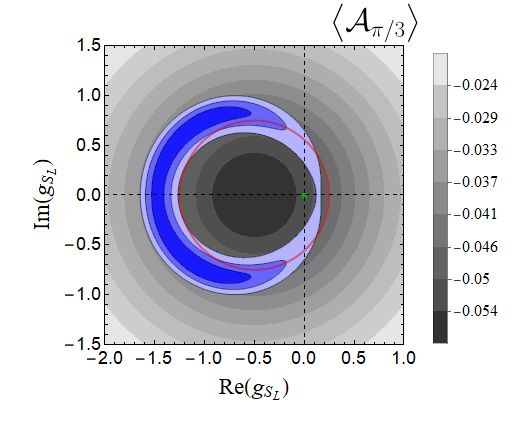}~\includegraphics[scale=.5]{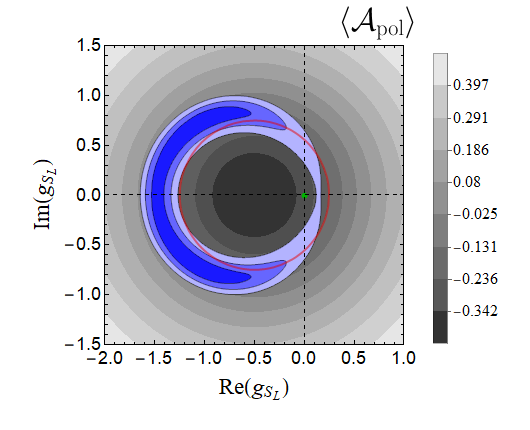}
        \caption{\sl  \small Scenario with only $g_{S_L}\neq 0$ shows that the only $\re ( g_{S_L}) $ and $\im ( g_{S_L}) $, allowed by $R_D^\mathrm{exp}$ and $R_{D^*}^\mathrm{exp}$, are complex and large (blue regions).  The gray regions correspond to various values of $\langle \mathcal{A}_{\pi/3}\rangle$ and to 
        $\langle \mathcal{A}_{\tau}\rangle$ in the left and right panel respectively. Both quantities are smaller than their SM counterparts, denoted by green dots in the plots.  Red circles correspond to $g_{S_L}$ consistent with the current $R_{\Lambda_c}^\mathrm{exp}$. }
     \label{fig:gSL}
\end{figure}

Another possibility is to only allow $g_T\neq 0$. Such a scenario could be built up from the scalar leptoquarks $R_2 =(3,2,7/6)$ and $S_1=(\bar 3,1,1/3)$, coupled in such a way that their respective non-zero $g_{S_L}$ cancel, in which case only $g_T\neq 0$ would survive.~\footnote{Note that the leptoquarks are specified by their SM gauge group quantum numbers.} In such a scenario, from the requirement of compatibility with $R_D^\mathrm{exp}$ and $R_{D^*}^\mathrm{exp}$, we again obtain a possibility of $\im(g_T)\neq 0$, which could be verified by measuring  $\langle D_4\rangle$ or $\langle E_4^\mathcal{A}\rangle$. Otherwise, $\langle \mathcal{A}_\mathrm{fb}\rangle$ appears to be more sensitive to this scenario than the other observables mentioned in this Section. 

\begin{figure}[h!]
\hspace*{-14mm} \includegraphics[scale=.5]{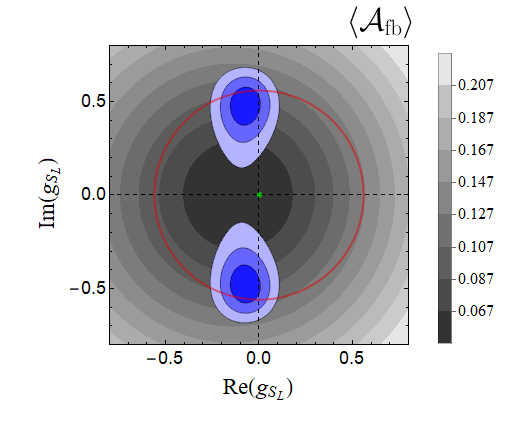}~\includegraphics[scale=.5]{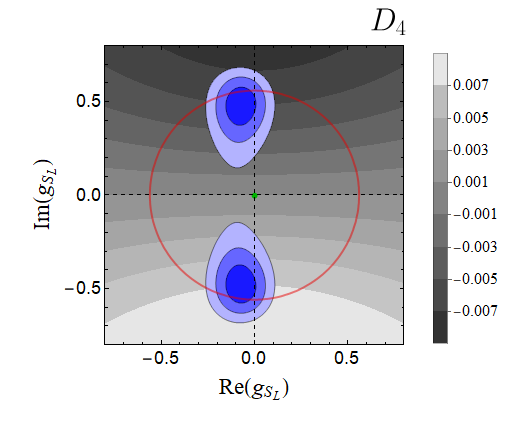}
        \caption{\sl \small  Scenario with non-zero NP couplings satisfying $g_{S_L}= 4 g_T$ is shown in the $\re ( g_{S_L}) $-$\im ( g_{S_L}) $ plane. The blue regions are selected by $R_D^\mathrm{exp}$ and $R_{D^*}^\mathrm{exp}$ to $1$, $2$ and $3\ \sigma$.  The gray regions correspond to various values of $\langle \mathcal{A}_\mathrm{fb}\rangle$ and to 
        $\langle D_4\rangle$ in the left and right panel respectively. Their SM values are indicated by green dots while the red circles limit the domain of $g_{S_L}$ consistent with $R_{\Lambda_c}^\mathrm{exp}$. }
     \label{fig:gSLgTplus}
\end{figure}

We now consider the scenarios that are often invoked when trying to accommodate $R_D^\mathrm{exp}$ and $R_{D^*}^\mathrm{exp}$ in a minimalistic NP scenario. In the first one, it is assumed that the $\mathcal{O}(1\,\mathrm{TeV})$ scalar leptoquark $R_2$~\cite{R2S3} is present. In that situation, at the high energy scale, we have $g_{S_L}= 4 g_T$ that originates when from the application of the Fierz identities when matching to the effective theory~\eqref{eq:lagrangian-lep-semilep}. That relation becomes $ g_{S_L}\simeq  8.1  g_T$ at the $\mu=m_b$, to which we refer in the following. A peculiarity of this scenario is that the $g_{S_L}$ resulting from compatibility with $R_D^\mathrm{exp}$ and $R_{D^*}^\mathrm{exp}$ has a large imaginary part, and different from zero to more than $3\sigma$. In order to check for the validity of this scenario, it is therefore of major importance to get $\langle D_4\rangle$ or $\langle E_4\rangle$. Other quantities are also important to measure since all of them, $\langle \mathcal{A}_\mathrm{fb}\rangle$, $\langle \mathcal{A}_{\pi/3}\rangle$, $\langle \mathcal{A}_{\tau}\rangle$, are likely to be smaller than their respective SM values. In Fig.~\ref{fig:gSLgTplus} we illustrate the situation for  $\langle \mathcal{A}_\mathrm{fb}\rangle$ and for $\langle D_4\rangle$. Since the sign of the phase is not constrained, the available values of $\langle D_4\rangle$ are symmetric.

\begin{figure}[h!]
\hspace*{-14mm} \includegraphics[scale=.5]{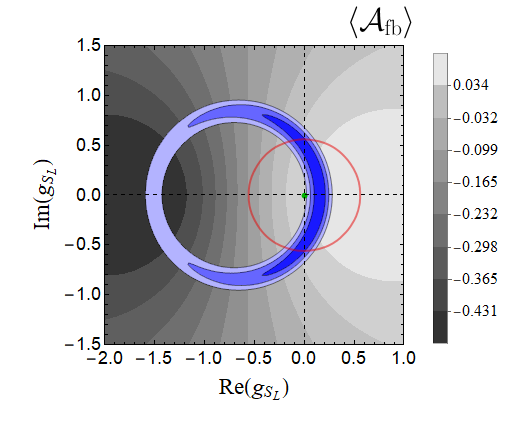}~\includegraphics[scale=.5]{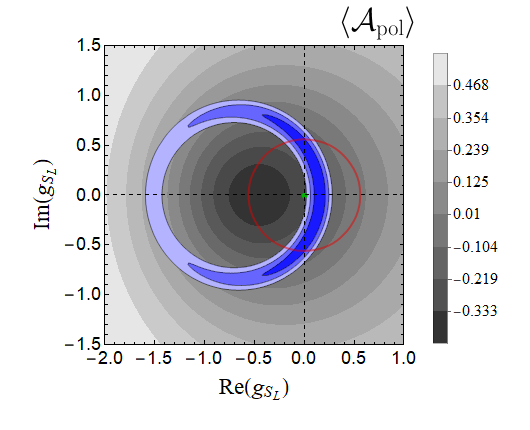}
        \caption{\sl \small  Scenario with non-zero NP couplings satisfying $g_{S_L}= - 4 g_T$. The values of $\re ( g_{S_L}) $-$\im ( g_{S_L}) $ in the blue regions are selected by $R_D^\mathrm{exp}$ and $R_{D^*}^\mathrm{exp}$ to $1$, $2$ and $3\ \sigma$.  The gray regions correspond to various values of $\langle \mathcal{A}_\mathrm{fb}\rangle$ and to 
        $\langle \mathcal{A}_\tau \rangle$ in the left and right panel respectively. The SM values are indicated by green dots while the red circles limit the domain consistent with $R_{\Lambda_c}^\mathrm{exp}$. }
     \label{fig:gSLgTminus}
\end{figure}

The second scenario in which $g_{S_L,T}\neq 0$, and which can explain $R_D^\mathrm{exp}$ and $R_{D^*}^\mathrm{exp}$ is the minimal extension of the SM by a low energy $S_1$ scalar leptoquark~\cite{S1S3}. In that scenario the two non-zero effective couplings are related via $g_{S_L}= - 4 g_T$, a relation that at the low energy scale $\mu=m_b$ becomes  $g_{S_L}\eqsim - 8.5 g_T$. Like in the previous cases, one would obviously prefer all the observables to be measured. None of the observables exhibits more pronounced sensitivity with respect to the others. We select to show in Fig.~\ref{fig:gSLgTminus} how $\langle \mathcal{A}_\mathrm{fb}\rangle$ and $\langle \mathcal{A}_\tau\rangle$ vary with respect to the SM when $g_{S_L}\neq 0$.

\subsection{More comments on the $g_{S_L}= + 4 g_T$ scenario}

\begin{figure}[h]
\centering \hspace*{-14mm} \raisebox{-0.5\height}{\includegraphics[scale=.66]{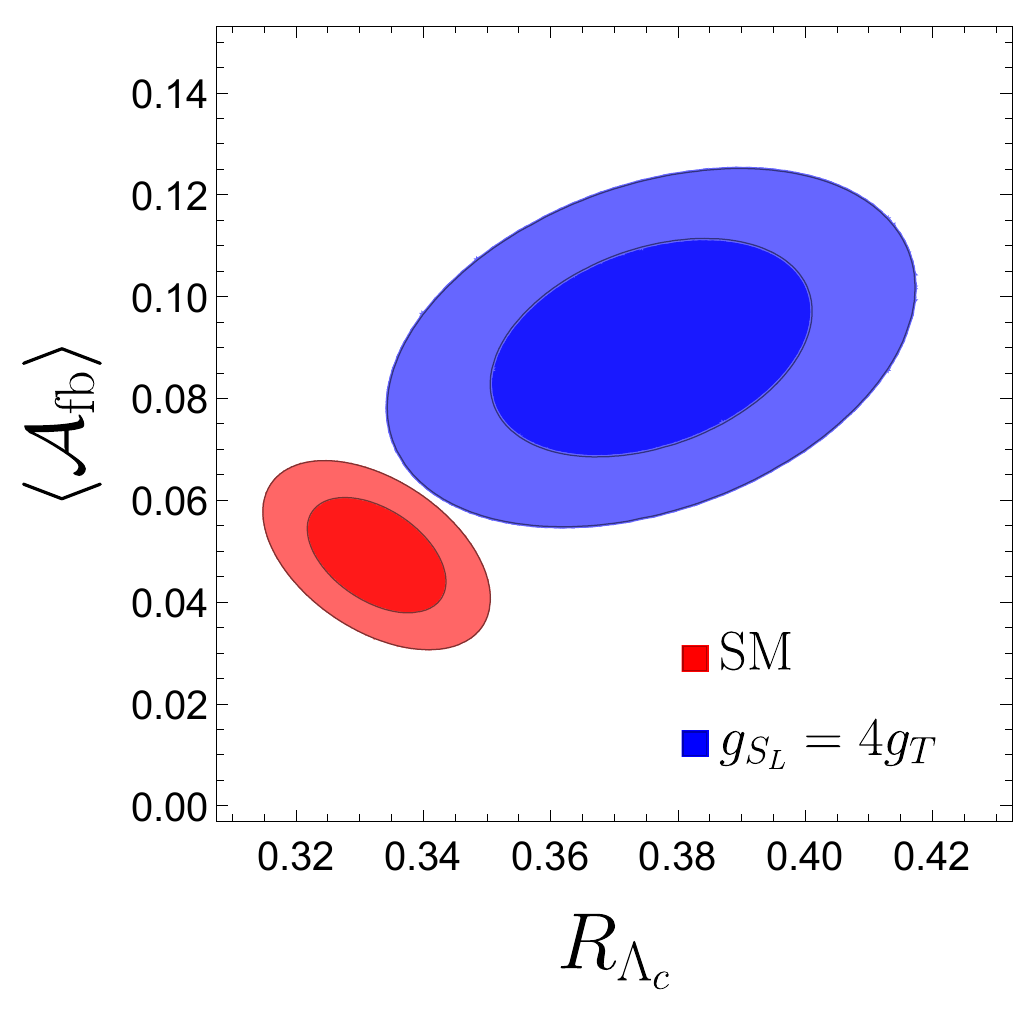}}~\qquad~\raisebox{-0.5\height}{\includegraphics[scale=.72]{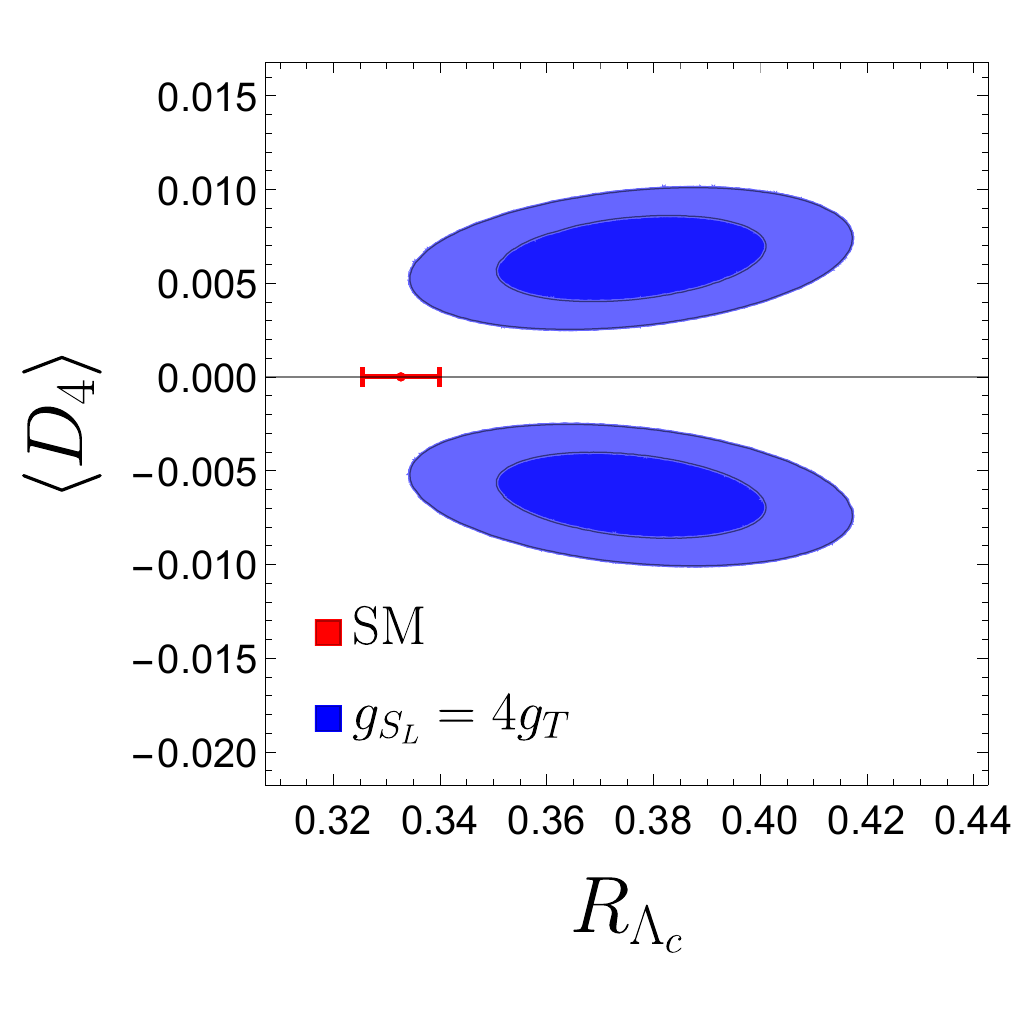}}
        \caption{\sl \small  Scenario of NP with a low energy scalar leptoquark $R_2$ verifying $g_{S_L}= + 4 g_T$. We show the regions of predicted values of 
         $R_{\Lambda_c}$, $\langle \mathcal{A}_\mathrm{fb}\rangle$ and $\langle D_4\rangle$, which clearly differ from SM, so that the measurement of these quantities can 
         help (in)validating this model. Note that the couplings are selected in such a way as to ensure the compatibility with $R_D^\mathrm{exp}$ and $R_{D^*}^\mathrm{exp}$ to $1$  and $2\ \sigma$. }
     \label{fig:corrs}
\end{figure}

As we showed above, the model in which the SM is extended by a presence of the $\mathcal{O}(1\,\mathrm{TeV})$ scalar leptoquark $R_2$ is peculiar because the 
compatibility with the measured $R_D^\mathrm{exp}$ and $R_{D^*}^\mathrm{exp}$ necessitates the NP coupling to have a large imaginary part. We show in Fig.~\ref{fig:corrs} how the measurement of three quantities, $R_{\Lambda_c}$, $\langle \mathcal{A}_\mathrm{fb}\rangle$ and $\langle D_4\rangle$, can help distinguishing  this scenario from the SM.  

\begin{figure}[h]
\centering \hspace*{-14mm} \includegraphics[scale=.45]{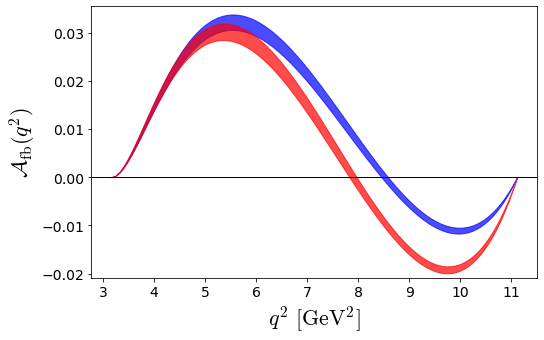}~\includegraphics[scale=.45]{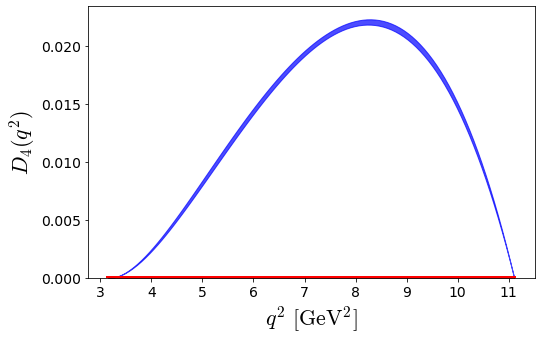}
        \caption{\sl \small  Displayed are the $q^2$-shapes of $\mathcal{A}_\mathrm{fb}(q^2)$ and $D_4(q^2)$, in the SM (red curves) and in the scenario with $g_{S_L}= 4 g_T$ where we choose the purely imaginary, $\im(g_{S_L})= 0.48$ (blue curves).  Notice that the zero of $\mathcal{A}_\mathrm{fb}(q^2)$ is larger in this NP scenario than in the SM. }
     \label{fig:shapes}
\end{figure}

Another point, which was abundantly discussed in the literature regarding $B\to K^\ast \ell^+\ell^-$, is the interest in defining the forward-backward asymmetry in one half of the available $q^2$-region. The situation with several observables, including the forward-backward asymmetry, is that they change the sign when moving from the low to high $q^2$'s.~\footnote{To be more specific, we find that $ \mathcal{A}_\mathrm{fb}(q^2)$, $B_2(q^2)$, $ \mathcal{A}_\tau (q^2)$ are the observables which change the sign when going from low to high $q^2$'s.}  As a result their integrated characteristics, cf. Eq.~\eqref{eq:intchar}, become small due to significant cancellations. 
It would therefore be beneficiary for this research if one could split the data to high and low $q^2$-regions, so that the absolute values of the resulting 
observables become larger. Note also that the shape of $D_4(q^2)$ is somewhat skewed towards the larger $q^2$'s when $\im(g_{S_L})\neq 0$, cf. Fig.~\ref{fig:shapes}. 

Finally, looking for the point $q_0^2$ at which a given observable changes the sign could provide us with a helpful information as well.
In particular, in the SM, we find that, 
\bea
\mathcal{A}_\mathrm{fb}^\mathrm{SM}(q^2_0) = 0, \quad \text{for} \quad q^2_0=  8.0(1)~\mathrm{GeV}^2\,.
\eea 
However, when switching on $\im(g_{S_L})= 0.48$, which is consistent with the scenario discussed in this subsection, that zero is shifted to a larger $q^2_0=  8.6(1)~\mathrm{GeV}^2$, see Fig.~\ref{fig:shapes}. 
 
\subsection{The case of $\Lambda_c\to \Lambda l   \nu$}

The expressions presented in Secs.~\ref{sec:eft} and \ref{sec:distr} are trivially transposable to other semileptonic decays of ground state baryons, $J^P=(1/2)^+$. 
When it comes to the phenomenological applications the main problem is that the relevant form factors have not yet been computed by means of lattice QCD. 
An important exception is $\Lambda_c\to \Lambda l   \nu$ for 
which all but the tensor form factors have been computed on the lattice~\cite{Meinel:2016dqj}. Since there is no phase space for a $\tau$-lepton in the final state, 
 experimentally accessible is the ratio
\bea
R_\Lambda^{(\mu/e)}  = \dfrac{\mathcal{B}(\Lambda_c\to \Lambda  \mu {\nu})}{\mathcal{B}(\Lambda_c\to \Lambda e  {\nu} )} , 
\eea
which has been recently measured to be $R_\Lambda^{(\mu/e)\,\mathrm{exp}}= 0.96\pm0.16\pm0.04$~\cite{BESIII:2016ffj}, and agrees with what we obtain after combining 
our expressions and the form factors from Ref.~\cite{Meinel:2016dqj}, namely, 
\bea
R_\Lambda^{(\mu/e)\,\mathrm{SM}}= 0.974(1)\,.
\eea
We can now assume that the NP modifies the coupling to $\mu \bar \nu$, but not to $e\bar \nu$, in the final state and express $R_\Lambda^{(\mu/e)}$ in a generic NP scenario as
\begin{equation}
\setlength{\jot}{12pt}
\begin{aligned}
\label{eq:Rlambda2}
R_{\Lambda}^{(\mu/e)} =  &\quad b_\mrm{ S}\, |g_S|^2 + b_\mrm{ VS}\, {\re\left[(1+g_{\mrm V}) {g_{\mrm S}}^\ast\right]} + b_\mrm{ P}\, |g_P|^2  + b_\mrm{ AP} \, \re\left[(g_A-1) {g_P}^\ast\right]   \nonumber \\
& + b_\mrm{ V}  \,  |1+g_V|^2 + b_\mrm{ A}\, |1-g_A|^2  \, .
\end{aligned}
\end{equation}
This formula is similar to Eq.~\eqref{eq:Rlambda} except that we neglect couplings to the tensor density because the corresponding form factors have not been computed on the lattice. 
The values of the coefficients $b_i$ in the above formula are:
\begin{align}
& b_\mrm{ S}=0.3777(110), & b_\mrm{ VS} =0.1312(41),  &\qquad\qquad b_\mrm{ P} =0.0696(21),  \nonumber \\
& b_\mrm{AP}=0.0391(12), & b_\mrm{ V} =0.2995(86), &\qquad\qquad b_\mrm{ A} =0.6743(90), 
\end{align}
and the associated correlation matrix reads
\[ 
\mrm{Corr}_{b} =
 \left(
\begin{array}{ccccccccc}
$1$ &	$0.8710$	& $0.0915$	& $0.0064$ & $0.1749$ & $-0.0843$\\
\cdot & $1$	& $-0.0529$ & $-0.0589$ & $0.3924$ & $-0.2997$\\
\cdot & \cdot & $1$	& $0.8925$	& $-0.0688$ & $0.1195$\\
\cdot & \cdot & \cdot & $1$	& $-0.0622$ & $0.1031$\\
\cdot & \cdot & \cdot & \cdot & $1$	& $-0.9910$\\
\cdot & \cdot & \cdot & \cdot & \cdot & $1$\\
\end{array}
\right)  \;,
\]
organized in a manner analogous to $\mrm{Corr}_a$ in Sec.~\ref{RLc-pheno}. 
In Ref.~\cite{Becirevic:2020rzi} the values of the NP couplings for the $c\to s\mu \nu$ based decays were constrained from the comparison of the measured $\mathcal{B}(D_s\to  l \bar \nu)$ and $\mathcal{B}(D\to K l \bar \nu)$ and the predictions obtained by using the hadronic matrix elements computed by means of lattice QCD. 
Due to smallness of such deviations and of the muon mass, we find that the angular observables are practically indistinguishable with respect to the SM values, cf. Tab.~\ref{tab:0}. 
%%%%%%%%%%%%%%%%%%%%%
\begin{table}[t]
\def\arraystretch{1.8}
\centering
\begin{tabular}{|c|c|c|c|c|c|}  \hline
Model & $R_{\Lambda}^{(\mu/e)}$ &  $\langle \mathcal{A}_{\mathrm{fb}}\rangle$ & $\langle\mathcal{A}_{\pi/3}\rangle$  & $\langle\mathcal{A}_{\mu}\rangle$ & $\langle\mathcal{A}_{\Lambda}\rangle$
   \\ \hline
SM        & $0.974(1)$  & $0.167(5)$    & $-0.125(4)$ &    $-0.910(2)$   &$-0.821(7)$      \\ \hline
{$g_{V_L}=0.06$}       & $1.094(1)$  & $0.167(5)$    & $-0.125(4)$ &    $-0.910(2)$   &$-0.821(7)$      \\ \hline
{$g_{S_L}=-0.03$}       & $0.969(1)$  & $0.170(5)$    & $-0.125(4)$ &    $-0.910(2)$   &$-0.821(7)$      \\ \hline
\end{tabular}
\caption{ \sl \small Best fit values for the couplings $g_{V_L}$ and $g_{S_L}$ have been obtained in Ref.~\cite{Becirevic:2020rzi} by assuming that the NP can affect the corresponding meson decays with a muon in the final state, and not an electron. We see that only $R_{\Lambda}^{(\mu/e)}$ is sensitive to the NP coupling either in a scenario in which only $g_{V_L}\neq 0$ or the one with $g_{S_L}\neq 0$. The other angular observables discussed in this Section, such as the forward-backward asymmetry $\langle \mathcal{A}_{\mathrm{fb}}\rangle$, muon polarization asymmetry $\langle\mathcal{A}_{\mu}\rangle$,  $\Lambda$-polarization asymmetry $\langle\mathcal{A}_{\Lambda}\rangle$ and the so-called ``$\pi/3$" asymmetry $\langle\mathcal{A}_{\pi/3}\rangle$, remain unchanged with respect to their respective SM values. }
\label{tab:0}
\end{table}
%%%%%%%%%%%%%%%%%%%%
Finally, and before closing this Section, we should also mention that a similar research can be extended to the case of $s\to u$ transitions~\cite{Chang:2014iba}. In particular, for the $\Lambda \to p l\bar \nu$ decay, BESIII recently reported $R_\Lambda^{(\mu/e)}= 0.178(28)$~\cite{BESIII:2021ynj}. However, no lattice QCD study of the relevant form factors has been made yet and the authors then have to  rely on the flavor SU(3) symmetry, combined with the results available in Refs.~\cite{Aoki:2019cca,Gupta:2018qil}.

\section{Summary and conclusion \label{sec:concl}}
\label{sec:conclusions}

In this work we revisited the problem of distinguishing the NP scenario in the exclusive $b \to c\ell \bar \nu$ decays, by focusing on the 
$\Lambda_b \longrightarrow \Lambda_c \ell \nu $ decay mode. This mode has received considerable attention in recent years. It has been studied at the LHCb, and more importantly the hadronic matrix elements relevant to the SM operators and those arising in the BSM scenarios have all been computed by means of numerical simulations of QCD on the lattice. 

By working in a general low energy effective theory, in which we included all of the possible NP contributions (without considering the right-handed neutrinos), we provided the expression 
for the angular distribution of this decay. In doing so we separated the contributions arising from various polarization states of the outgoing baryon and lepton.  
In that way we were able to show that one can build at most $10$ different observables. That number rises to $18$, if one considers the secondary decay, which we choose to be $\Lambda_c \to \Lambda \pi$. Notice that we can have extra $18$ observables if we also included the coefficients that would come with the polarization asymmetries 
regarding the final $\Lambda$. However, those extra observables are not informative as far as NP is concerned but they would lead to yet another determination of the polarization asymmetry parameter of $\Lambda$, referred to in the literature as $\alpha$. 

In an ideal scenario, one would prefer to measure as many observables as possible in order to test the viability of various scenarios of physics BSM. In our phenomenological analysis we restrained our attention to a subset of $6$ observables which we find to exhibit more pronounced sensitivity to the non-zero NP couplings.  In simplified scenarios, used to accommodate $R_{D^{(\ast)}}^\mathrm{exp}$, one is turning on one coupling at the time.

If we assume NP to arise from the $g_{V_L}$ coupling, we find that all our observables remain SM-like, except for the ratio $R_{\Lambda_c} = |1+g_{V_L}|^2 R_{\Lambda_c}^\mathrm{SM}$. 

Other simplified scenarios include an $\mathcal{O}(1\,\mathrm{TeV})$ scalar leptoquark, giving rise to two NP couplings, $g_{S_L}$ and $g_T$, couplings to the (left) scalar and tensor quark operators respectively. Due to Fierz identities these couplings are  related to each other  as   
$g_{S_L}=\pm 4 g_T$, at the high energy scale. In these scenarios all observables can be very different from their SM counterparts. We isolated a few such observables to show how they can be used to validate or refute the scenarios that are currently used in order to describe the deviations of $R_{D^{(\ast)}}^\mathrm{exp}$ from their SM values, $R_{D^{(\ast)}}^\mathrm{SM}$. In particular, to accommodate such discrepancies in the scenario with $g_{S_L}=+ 4 g_T$, the NP coupling must have a non-zero complex phase, in which case some of the observables (such as $\langle D_4\rangle$) would be a clear test of validity of such a scenario because one can have $\langle D_4\rangle \neq 0$ only if $\im ( g_{S_L})\neq 0$. 

We also discuss the impact of the recently reported $R_{\Lambda_c}^\mathrm{exp}$, the result which can and should be experimentally improved. Importantly, however, we must emphasize that the observables arising from the angular distribution, such as those discussed in this paper, represent a fine and powerful check of presence of NP at low energy scales. One can, for example, easily build a scenario in which $R_{D^{(\ast)}}$, $R_{J/\psi}$ and $R_{\Lambda_c}$ are consistent with their SM values but with several of the angular observables considerably different from their SM predictions. It is therefore important to measure these pbservables. Throughout our phenomenological discussion we referred to the quantities integrated over the available $q^2$'s. Needless to say that in some cases, such as the forward-backward asymmetry with respect to the outgoing lepton, the $q^2$-dependence of the observables could provide us with a very interesting information and potentially reveal the presence of physics BSM. 

We should also stress that the hadronic form factors for all of the operators needed for the full NP analysis of this decay have been computed on the lattice, which is not the case with the modes involving mesons, such as $B\to D^{(\ast)}\ell\bar \nu$, for which the tensor form factors have not been computed on the lattice. It should be kept in mind, however, that the hadronic matrix elements relevant to $\Lambda_b \to \Lambda_c \ell\bar \nu$ have been computed by only one lattice group and it is of major importance for this research that another lattice QCD study is made, preferably by using a different discretization of QCD. 

The above analysis is easily applicable to other semileptonic decays of ground state baryons.  showed that in the case of $\Lambda_c\to \Lambda l\nu$ the angular observables 
\vskip 2cm 
\section*{Acknowledgments}
%%%%%%%%%%%%%%%%%%%%%%%%%%%%%%%%%%%%%%%%%%%%%%%%%%
This work has been supported in part by the European Union's Horizon 2020 research and innovation programme under the Marie Sklodowska-Curie grant agreement N$^\circ$~660881-Hidden.
%%%%%%%%%%%%%%%%%%%%%%%%%%%%%%%%%%%%%%%%%%%%%%%%%%

\newpage 

\section*{Appendix}
%%%%%%%%%%%%%%%%%%%%%%%%%%%%%%%%%%%%%%%%%%%%%%%%%%
In this appendix we give additional details which might be important for a reader willing to repeat the computation the results of which are presented in the body of this paper. 
To discuss the kinematics of $\Lambda_b(p)\to \Lambda_c(k) \ell(k_1) \bar\nu(k_2)$, we introduce $q  =  k_1+k_2 = p-k$, and choose a $z$-axis along the flight of  $\Lambda_c$. Angle $\theta$ is defined in the frame in which $|\vec q |=0$ between the $z$-axis and the direction of flight of $\ell$.

In the $\Lambda_b$ rest frame we then have: $M_{\Lambda_b}= E_{\Lambda_c} +q_0$. By combining $M_{\Lambda_b}^2 = M_{\Lambda_c}^2+q^2+2k\cdot q$ with $k\cdot q  = E_{\Lambda_c}q_0+q_z^2 =M_{\Lambda_b}q_0-q^2$ we get
\begin{align}
q_0 &= \frac{M_{\Lambda_b}^2-M_{\Lambda_c}^2+q^2}{2M_{\Lambda_b}},
\quad E_{\Lambda_c} = \frac{M_{\Lambda_b}^2+M_{\Lambda_c}^2-q^2}{2M_{\Lambda_b}},
\quad q_z = \sqrt{q_0^2-q^2} = \frac{\sqrt{\lambda_{{\Lambda_b}{\Lambda_c}}(q^2)}}{2M_{\Lambda_b}},
\end{align}
where we use 
\begin{align}
\lambda_{{\Lambda_b}{\Lambda_c}}(q^2) &= M_{\Lambda_b}^4+M_{\Lambda_c}^4+q^4-2M_{\Lambda_b}^2M_{\Lambda_c}^2-2M_{\Lambda_b}^2q^2-2M_{\Lambda_c}^2q^2 = Q_+Q_-,
\end{align}
where 
\begin{equation}
Q_\pm = \left(M_{\Lambda_b}\pm M_{\Lambda_c}\right)^2-q^2\, , 
\end{equation}
which we already used in Sec.~\ref{sec:eft}. 

In the dilepton rest frame:
\begin{align}
\begin{pmatrix}
\sqrt{q^2}\\
0\\
0\\
0
\end{pmatrix}&=\begin{pmatrix}
E_\nu\\
-p_\ell\sin \theta \\
0\\
-p_\ell\cos \theta
\end{pmatrix}+\begin{pmatrix}
E_\ell\\
p_\ell\sin \theta\\
0\\
p_\ell\cos \theta
\end{pmatrix},
\end{align}
and
\begin{align}
k_1\cdot k_2 &= \frac{q^2-m_\ell^2}{2},\quad E_\ell = \frac{q^2+m_\ell^2}{2\sqrt{q^2}},\quad 
E_\nu  = \frac{q^2-m_\ell^2}{2\sqrt{q^2}}.
\end{align}
To go from the first frame to the other, we use a Lorentz boost, 
$\sqrt{q^2} =\gamma q_0+\beta\gamma q_z$, $0 = \beta\gamma q_0 +\gamma q_z$, so that the boost parameters are $\beta = -q_z/q_0$
and $\gamma = q_0/\sqrt{q^2}$, and therefore the components of the momenta of hadrons in the second frame are:
\begin{align}
p=& \frac{1}{ \sqrt{q^2}}  \left( M_{\Lambda_b} q_0  ,0,0,M_{\Lambda_b} q_z \right) = \frac{1}{2\sqrt{q^2}} \left( M_{\Lambda_b}^2-M_{\Lambda_c}^2+q^2, 0 , 0 , \sqrt{\lambda_{{\Lambda_b}{\Lambda_c}}(q^2)} \right),\\
k= &\frac{1}{ \sqrt{q^2}} \left( q_0 (M_{\Lambda_b} - q_0)+q_z^2 ,0,0,q_z^2\right) = \frac{1}{2\sqrt{q^2}} \left( M_{\Lambda_b}^2-M_{\Lambda_c}^2- q^2, 0 , 0 , \sqrt{\lambda_{{\Lambda_b}{\Lambda_c}}(q^2)} \right),
\end{align}
from which one can compute other scalar products.

For completeness, we also give the expression for the spinors in the Dirac basis:
\begin{align}
u(\lambda_\ell=\pm 1/2 ) &= \sqrt{E_\ell+m_\ell}\begin{pmatrix}
\xi_\pm\\
\frac{\vec \sigma\cdot \vec p_\ell}{E_\ell+m_\ell}\xi_\pm
\end{pmatrix}, \quad 
v(\lambda_\ell=\pm 1/2 ) = \sqrt{E_\ell+m_\ell}\begin{pmatrix}
\frac{\vec \sigma \cdot \vec p_\ell}{E_\ell+m_\ell}\xi_\mp\\
\xi_\mp
\end{pmatrix},
\end{align}
where $\vec \sigma$ are the Pauli matrices and the spinors $\xi_\pm$  are given by
\begin{align}
\xi_+ &= \exp\left(i\frac{\vec \sigma\cdot \vec \theta}{2}\right)\begin{pmatrix}
1\\
0
\end{pmatrix},\qquad
\xi_-  =  \exp\left(i\frac{\vec \sigma \cdot\vec \theta}{2}\right)\begin{pmatrix}
0\\
1
\end{pmatrix}.
\end{align}
Explicitly, for the lepton in the dilepton rest-frame, we have
\begin{align}
\xi_+ &= \begin{pmatrix}
\cos \frac{\theta}{2} \\
\sin \frac{\theta}{2}\\
\end{pmatrix}, \qquad
\xi_- = \begin{pmatrix}
-\sin \frac{\theta}{2} \\
\cos \frac{\theta}{2} \\
\end{pmatrix},
\end{align}
and
\begin{align}
u_{\ell,+1/2} &= \sqrt{E_\ell+m_\ell}\begin{pmatrix}
\cos \frac{\theta}{2} \\
\sin \frac{\theta}{2} \\
\frac{p_\ell}{E_l+m_\ell}\cos \frac{\theta}{2} \\
\frac{p_\ell}{E_\ell+m_\ell} \sin \frac{\theta}{2} 
\end{pmatrix},\qquad
u_{\ell, -1/2}  = \sqrt{E_\ell+m_\ell}\begin{pmatrix}
-\sin\frac{\theta}{2}\\
\cos\frac{\theta}{2}\\
\frac{p_\ell}{E_\ell+m_\ell}\sin \frac{\theta}{2}\\
-\frac{p_\ell}{E_\ell+m_\ell}\cos \frac{\theta}{2}
\end{pmatrix}.
\end{align}
For the neutrino in the dilepton rest-frame, we take $\theta \to \theta+\pi$:
\begin{align}
v_{\nu, +1/2} &= \sqrt{E_\nu}\begin{pmatrix}
\cos\frac{\theta}{2}\\
\sin\frac{\theta}{2}\\
-\cos \frac{\theta}{2} \\
-\sin \frac{\theta}{2} 
\end{pmatrix}.
\end{align}
As for the baryons, in the $\Lambda_b$ rest frame, $|\vec \theta |=0$, we simply have
\begin{align}
u_{\Lambda_b, +1/2} &= \sqrt{2M_{\Lambda_b}}\begin{pmatrix}
1\\
0\\
0\\
0
\end{pmatrix},\quad
u_{\Lambda_b, -1/2} = \sqrt{2M_{\Lambda_b}}\begin{pmatrix}
0\\
1\\
0\\
0
\end{pmatrix},\nonumber \\
u_{\Lambda_c, +1/2}  &= \sqrt{M_{\Lambda_c}+E_{\Lambda_c}}\begin{pmatrix}
1\\
0\\
\frac{q_z}{M_{\Lambda_c}+E_{\Lambda_c}}\\
0
\end{pmatrix},\quad
u_{\Lambda_c, -1/2} = \sqrt{M_{\Lambda_c}+E_{\Lambda_c}}\begin{pmatrix}
0\\
1\\
0\\
\frac{-q_z}{M_{\Lambda_c}+E_{\Lambda_c}}
\end{pmatrix}.
\end{align}
Finally, the spinors for $\Lambda_c$ and $\Lambda$ in the $\Lambda_c$ rest frame, $\vec \theta  = \phi \vec e_z + \theta_\Lambda \vec e_x$, read:
\begin{align}
\xi_+ &= \begin{pmatrix}
e^{\frac{i\phi}{2}} \cos \frac{\theta_\Lambda}{2} \\
\sin \frac{\theta_\Lambda}{2}\\
\end{pmatrix}, \qquad
\xi_- = \begin{pmatrix}
-\sin  \frac{\theta_\Lambda}{2} \\
e^{\frac{-i\phi}{2}} \cos \frac{\theta_\Lambda}{2}\\
\end{pmatrix},
\end{align}
and
\begin{align}
u_{\Lambda_c, +1/2} &= \sqrt{2M_{\Lambda_c}}\begin{pmatrix}
1\\
0\\
0\\
0
\end{pmatrix},\qquad
u_{\Lambda_c, -1/2}  = \sqrt{2M_{\Lambda_c}}\begin{pmatrix}
0\\
1\\
0\\
0
\end{pmatrix},\nonumber\\
u_{\Lambda,+1/2} &= \sqrt{E_\Lambda+M_\Lambda}\begin{pmatrix}
\cos \frac{\theta_\Lambda}{2}\, e^{\frac{i\phi}{2}}\\
\sin \frac{\theta_\Lambda}{2} \\
\frac{p_\ell}{E_\Lambda+M_\Lambda}\cos \frac{\theta_\Lambda}{2}\, e^{\frac{i\phi}{2}}\\
\frac{p_\ell}{E_\Lambda+M_\Lambda}\sin \frac{\theta_\Lambda}{2} 
\end{pmatrix},\quad
u_{\Lambda, -1/2} = \sqrt{E_\Lambda+M_\Lambda}\begin{pmatrix}
-\sin \frac{\theta_\Lambda}{2} \\
\cos \frac{\theta_\Lambda}{2}\, e^{-\frac{i\phi}{2}}\\
\frac{p_\ell}{E_\Lambda+M_\Lambda}\sin \frac{\theta_\Lambda}{2} \\
-\frac{p_\ell}{E_\Lambda+M_\Lambda}\cos \frac{\theta_\Lambda}{2} \, e^{-\frac{i\phi}{2}}
\end{pmatrix}.
\end{align}

\end{document}